\documentclass[amsmath,amssymb,twocolumn,aps,prb,footinbib,superscriptaddress]{revtex4-1}

\usepackage{mathtools}
\usepackage{mathrsfs, amsmath, wasysym}
\usepackage[mathscr]{eucal}
\usepackage{xfrac}
\usepackage{graphicx}
\usepackage{dcolumn}
\usepackage{bm}
\usepackage{color}
\usepackage{tabularx}
\usepackage{natbib}
\usepackage[colorlinks,linkcolor=blue,citecolor=blue,urlcolor=blue]{hyperref}
\usepackage[normalem]{ulem}
\usepackage[all]{hypcap}
\usepackage[dvipsnames,usenames,table]{xcolor}

\newcommand{\nocontentsline}[3]{}
\newcommand{\tocless}[2]{\bgroup\let\addcontentsline=\nocontentsline#1{#2}\egroup}

\newcommand{\be}{\begin{equation}}
\newcommand{\ee}{\end{equation}}
\newcommand{\beg}{\begin{gather}}
\newcommand{\eeg}{\end{gather}}
\newcommand{\beq}{\begin{eqnarray}}
\newcommand{\eeq}{\end{eqnarray}}
\newcommand{\bea}{\begin{align}}
\newcommand{\eea}{\end{align}}
\newcommand{\beqq}{\begin{eqnarray*}}
\newcommand{\eeqq}{\end{eqnarray*}}

\renewcommand{\Im}{{\text{Im}\,}}
\renewcommand{\Re}{{\text{Re}\,}}

\begin{document}

\title{Effect of quasiparticles on the parameters of a gap-engineered transmon}

\author{Daniil S. Antonenko}
\affiliation{Department of Physics, Yale University, New Haven, Connecticut 06520, USA}

\author{Pavel D. Kurilovich}
\affiliation{Department of Physics, Yale University, New Haven, Connecticut 06520, USA}
\affiliation{Department of Applied Physics, Yale University, New Haven, Connecticut 06520, USA}

\author{Francisco J. Matute-Ca\~{n}adas} 
\affiliation{Departamento de F\'{ı}sica Te\'{o}rica de la Materia Condensada,
Condensed Matter Physics Center (IFIMAC) and Instituto Nicol\'{a}s Cabrera,
Universidad Aut\'{o}noma de Madrid, Madrid 28049, Spain}

\affiliation{Department of Physics, Yale University, New Haven, Connecticut 06520, USA}

\author{Leonid I. Glazman}
\affiliation{Department of Physics, Yale University, New Haven, Connecticut 06520, USA}
\affiliation{Yale Quantum Institute, Yale University, New Haven, Connecticut 06520, USA}

\begin{abstract}
We evaluate the quasiparticle contribution to the frequency shift and relaxation rates of a transmon with the Josephson junctions connecting superconductors that have unequal energy gaps. The gap difference substantially affects the transmon characteristics. We investigate their dependence on the density and effective temperature of the quasiparticles, and on the nominal (unperturbed by the quasiparticles) transmon frequency. 
At temperatures low compared to the qubit frequency, the gap difference can induce an anomalous positive frequency shift, resulting in a non-monotonic temperature dependence of the transmon frequency.
The qubit relaxation rate exhibits a resonance when the qubit frequency matches the gap difference; the shape of the resonance is strongly temperature-dependent.
We propose to use these effects to access the details of the quasiparticle energy distribution.

\end{abstract}

\maketitle

\section{Introduction}

Quasiparticles are ubiquitous in superconducting devices. In qubits, they affect key performance metrics, such as the $T_1$ and $T_2$ times. Low relaxation times, in turn, limit the fidelity of quantum operations. Although thermal pair breaking is expected to be negligible at millikelvin temperatures, numerous experiments confirm a persistent population of resident quasiparticles~\cite{CatelaniGlazman-qp-SciPost, Non-eq-qp-Devoret-2004, Qp-decay-rate-LutchynGlazmanLarkin, Kinetics-qubit-qp-LutchynGlazmanLarkin, Kinetics-noneq-qp-Echternach, EnergyDecayQp-MartinisAumentado, SparseQP-Klapwijk, CatelaniGlazman-qpRelaxFlux, MicrowaveQP-deVisserKlapwijk, Single-qp-trapping, riste_millisecond_2013, MeasurementQPDynamics-NatureComm2014, Hot-qp-transmon-Houzet-Devoret}. These excitations are produced by stray radiation~\cite{Serniak-PhotonAssisted, liu2024}, cosmic rays impacts~\cite{wilen_correlated_2021,mcewen_resolving_2022, harrington_synchronous_2024}, stress-release events~\cite{yelton2025}, or other poorly understood phenomena. The density of the resident quasiparticles is roughly constant in the temperature range where the thermal quasiparticle population is negligible. 
While the mere presence of a small quasiparticle density does not affect the qubit performance, it is their tunneling across the Josephson junction that degrades it. 
Recently, the gap engineering technique, aimed to impede the quasiparticle tunneling, has gained traction \cite{GapEngineering-Sun-PRL-2012, GapEngineering-RiwarCatelani2019, GapEngineering-NatureReview-2021, Catelani-asymmetric,GapEngineering-PanYuNature2022, mcewen_resisting_2024, Vlad-new-ErrorBursts,kishmar2025}. The simplest gap-engineered device is an asymmetric Josephson junction with different superconducting gaps in the leads. Practically, such junctions can be fabricated by leveraging the fact that the gap depends on the material thickness \cite{Chubov-Al, Yamamoto-Al, Court-Al}.
To suppress the adverse quasiparticle effects, the gap difference $\delta\Delta$ in the fabricated devices \cite{Vlad-new-ErrorBursts} is made substantially larger than the qubit frequency $\omega_{10}$ (hereinafter we set units with $\hbar=1$). In that case, a low-energy quasiparticle can not cross the junction even after receiving energy $\omega_{10}$ from the qubit.

Gap engineering is also used for studying the quasiparticles~\cite{Connolly-nonequilibrium-density, RecoveryDynamics-2025}. The details of their generation, dynamics, and energy distribution remain enigmatic so far, calling for further experimental and theoretical study. At low quasiparticle densities, an efficient experimental way to study them is by measuring their effect on qubit properties \cite{Catelani-qp, Connolly-nonequilibrium-density}. Using a qubit as a spectroscopic tool is facilitated by a split-transmon scheme incorporating gap-engieered devices\cite{Diamond-distinguishing-parity-switching}. This scheme allows tuning of the qubit frequency $\omega_{10}$, so that the energy difference $T_*=\delta\Delta-\omega_{10}$ in a given device can be varied by means of a magnetic flux. For a quasiparticle residing in the lead with the lower gap to cross the junction, the (positive) energy deficit $T_*$ should be covered by the quasiparticle energy.

\begin{figure}[h]
    \centering \includegraphics[width=0.95\linewidth]{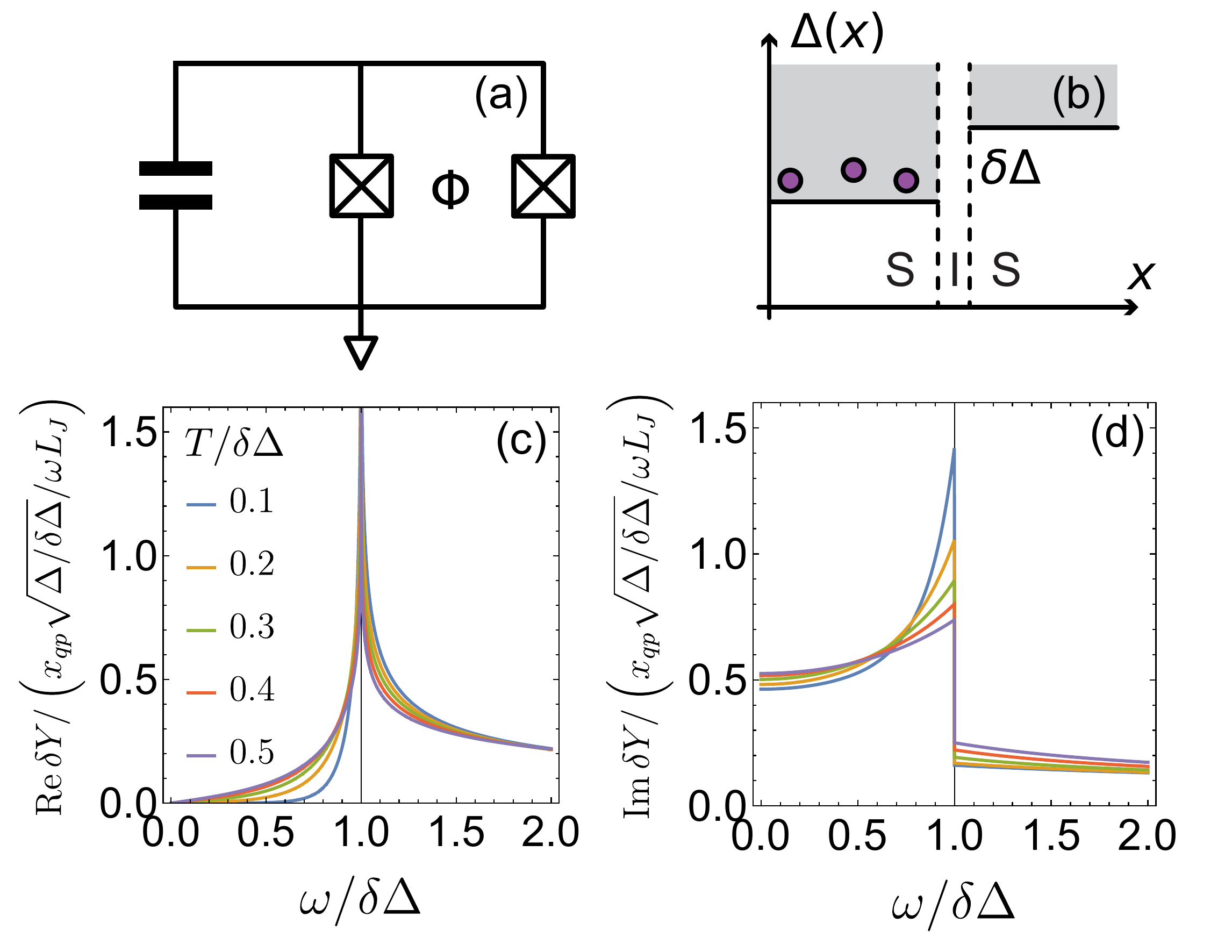}
    \caption{
    (a) Split-transmon effective circuit; (b) Josephson junction connecting leads with unequal superconducting gaps, $\delta\Delta = \Delta_R - \Delta_L$. At low energies, quasiparticles reside in the lead with the lower gap;  (c, d) Quasiparticle contribution to the  admittance of the junction as a function of the normalized frequency, $\omega/\delta\Delta$, at various values of $T / \delta \Delta$  and zero phase bias of the junction, $\varphi_0 = 0$. The two panels show the real (c) and imaginary (d) parts of the contribution. At any temperature, the former exhibits a logarithmic divergence and the latter a step at $\omega=\delta\Delta$. The limit $\mathcal{V}_L \gg \mathcal{V}_R$ was taken, rendering $\zeta(T) = 1$, see Eq.~\eqref{f0-via-xqp}. In all plots, $\delta\Delta/\Delta \ll 1$.
\label{fig:Y-one-asymm}
    }
\end{figure}

In this work, we focus primarily on the quasiparticle effects in devices with $|T_*|\ll\omega_{10}$. 
We investigate the quasiparticle-induced qubit frequency shift $\delta\omega$ and relaxation rates as a function of the effective quasiparticle temperature. The temperature dependence becomes especially strong and non-monotonic at $T_*\ll\omega_{10}$, making these characteristics a valuable resource for studying the low-energy domain of the quasiparticle distribution function. 

The knowledge of the distribution function is important. At a fixed density of non-equilibrium quasiparticles, the ``colder'' they are, the better is the qubit performance. The equilibration processes and distribution function of quasiparticles are being actively investigated now~\cite{Vlad-new-ErrorBursts,Connolly-nonequilibrium-density}. Theory predicts a very slow quasiparticle relaxation at low energies, but the specific form of the energy dependence of the relaxation rate is still debated \cite{CatelaniGlazman-qp-SciPost, SavichGlazmanKamenev, SkvortsovStepanov-GranularTheory}.
The ability to tune $T_*$ down to zero may help future experiments address the quasiparticle dynamics.

Our theory uses the standard method of tunnel Hamiltonian (Section~\ref{sec:model}). To address the transmon properties, we start by revisiting the evaluation of the admittance of a Josephson junction as a function of frequency and quasiparticle distribution function, focusing on the case of non-equal gaps in the junction leads (Sections~\ref{sec:dc-josephson} and \ref{sec:finiteomega}). We evaluate the quasiparticle-induced frequency shift in Section~\ref{sec:frequency-shift}, central in this paper. Section~\ref{sec:transition-rates} is devoted to the evaluation of charge-parity switching rates. 
We conclude in Section~\ref{sec:discussion} and provide additional technical details in the Appendix.

\section{The model}
\label{sec:model}

Coupling between the leads of the Josephson junction can be written in terms of the electrons creation and annihilation operators in each lead $c_{n_{L(R)},\sigma}$, $c_{n_{L(R)},\sigma}^\dagger$:
\be \label{H-tunnel}
	H_T = \sum_{n_{L},n_{R},\sigma}t_{n_{L}n_{R}}e^{i(\varphi_{L}-\varphi_{R})/2}c_{n_{L}\sigma}^{\dagger}c_{n_{R}\sigma}+\text{h.c.},
\ee
where $\varphi_{R(L)}$ is the superconducting phase in each lead, and $t_{n_{L}n_{R}}$ is tunneling amplitude of the spin-$\sigma$ electron from the level $n_{R}$ in the right terminal to the $n_{L}$ level in the left terminal. 

Here we assume that both leads can be described by the BCS theory of superconductivity. Electron operators are related to the quasiparticle creation operators by the Bogoliubov transform, which takes the following form: 
\be \label{Bogoliubov-transform}
	c_{n,\sigma} = u_{n}\gamma_{n,\sigma}+\sigma v_{n}\gamma_{n,-\sigma}^{\dagger},
\ee
with the electron and hole amplitudes
\be \label{uv-expressions}
	u_{n}=\sqrt{\frac{1}{2}\left(1+\frac{\xi_{n}}{\epsilon_{n}}\right)},\quad v_{n}=\sqrt{\frac{1}{2}\left(1-\frac{\xi_{n}}{\epsilon_{n}}\right)}, 
\ee
where $n$ stands for $n_{L(R)}$ level in each lead and $\xi_n$ is its energy in the normal state. The quasiparticle energies are given by
\be
	\epsilon_{n_{L(R)}} = \sqrt{ \xi_{n_{L(R)}}^2 + \Delta_{L(R)}^2 } .
\ee
We will assume that quasiparticles in both leads are described by the same distribution function that depends only on their energy,
\be
	f_{\epsilon} = \langle \gamma_{n_{L(R)},\sigma}^\dagger \gamma_{n_{L(R)},\sigma}  \rangle .
\ee
In practice, superconducting qubits and resonators have a small number of quasiparticles so that $f_{\epsilon} \ll 1$ for all $\epsilon$.

Superconducting gap $\Delta_{L(R)}$ in each lead should be determined self-consistently. Its dependence on the quasiparticle distribution is simplified when the temperature $T$ (or typical energy of non-equilibrium quasiparticles) is much smaller than $\Delta_{L(R)}$. For convenience, we introduce quasiparticle density 
\be \label{xqp-definition}
	x_{qp,L(R)} = \frac{n_{qp,L(R)}}{2 \nu_{L(R)} \Delta_{0,L(R)}}
\ee
normalized by superconducting gap in the absence of quasiparticles, $\Delta_{0,L(R)}$, and electron density of states per one spin projection, $\nu_{L(R)}$. 
In the limit  $x_{qp,L(R)} \ll 1$, we may use the perturbative expansion, \cite{OwenScalapino}
\be \label{gap-suppression}
    \Delta_{L(R)} = \Delta_{0,L(R)} (1 - x_{qp,L(R)}),
\ee
sufficient for the majority of relevant experiments. We will express the physical properties through $\Delta_{L(R)}$ below. The quasiparticle density in Eq.~\eqref{gap-suppression} includes both particles thermally generated at equilibrium, and the resident quasiparticles generated by external sources and persisting in practical cases down to the lowest temperatures~\cite{CatelaniGlazman-qp-SciPost, Non-eq-qp-Devoret-2004, Qp-decay-rate-LutchynGlazmanLarkin, Kinetics-qubit-qp-LutchynGlazmanLarkin, Kinetics-noneq-qp-Echternach, EnergyDecayQp-MartinisAumentado, SparseQP-Klapwijk, CatelaniGlazman-qpRelaxFlux, MicrowaveQP-deVisserKlapwijk, Single-qp-trapping, riste_millisecond_2013, MeasurementQPDynamics-NatureComm2014, Hot-qp-transmon-Houzet-Devoret}. 
Therefore, we will take the system-averaged density $x_{qp}$ [see Eq.~\eqref{xqp-through-nqp}] as a free parameter representing both thermal and resident quasiparticles.

\section{DC Josephson effect in asymmetric junction}
\label{sec:dc-josephson}

A well-known consequence of the tunnel coupling Eq.~\eqref{H-tunnel} is the dc Josephson current \cite{BaronePaterno}
\be \label{Josephson-dc}
	I_J = \frac{\hbar}{2e} \frac{1}{L_J^{\text{full}}} \sin \varphi_0
\ee
in the presence of a constant phase difference $\varphi_0$ between the leads. Here low-frequency kinetic inductance $L_J^{\text{full}}$ of the junction equals
\be \label{LJfull-expression}
	\frac{1}{L_J^{\text{full}}} = \frac{1}{L_J} + \frac{1}{L_J^{qp}},
\ee
where
\begin{align}
\label{LJ-expression}
	\frac{1}{L_J} & = 8 e^2 \sum_{n_{L}n_{R}\sigma}  \bar{t^2} \, u_{L}u_{R}v_{L}v_{R} \frac{1}{\epsilon_{L}+\epsilon_{R}},
\\
\label{LJqp-expression}
	\frac{1}{L_J^{\text{qp}}} & = 8 e^2 \sum_{n_{L}n_{R}\sigma}  \bar{t^2} \, u_{L}u_{R}v_{L}v_{R} \Big(\frac{f_{L}-f_{R}}{\epsilon_{L}-\epsilon_{R}} - \frac{f_{L} + f_{R}}{\epsilon_{L}+\epsilon_{R}} \Big).
\end{align}
Here $u_{L(R)}$, $v_{L(R)}$, $f_{L(R)}$ stand for $u_{n_{L(R)}}$, $v_{n_{L(R)}}$, $f_{\epsilon_{L(R)}}$. Averaging in $\bar{t^2}$ is assumed to be made over many levels $n$, yielding its relation to the normal-state conductance, $G_N = 4\pi e^2 \nu_L \nu_R \bar{t^2}$.
The dependence of $1/L_J$ on $x_{qp}$ comes only through the dependence of $\Delta$ on $x_{qp}$, see Eq.~(\ref{gap-suppression}). Term $1/L_J^{\text{qp}}$ depends on $x_{qp}$ through the distribution functions $f_L$, $f_R$; to the linear order in $x_{qp}$ we may set here $\Delta=\Delta_0$.

Evaluation of Eq.~(\ref{LJ-expression}) results in
\be \label{LJ-asymmetric}
	\frac{1}{L_J} = 4 G_N \frac{ \Delta_L \Delta_R}{\Delta_L + \Delta_R} K\left( \frac{|\Delta_L - \Delta_R|}{\Delta_L + \Delta_R} \right),
\ee
where $K$ is the complete elliptic integral of the first kind. Eq.~\eqref{LJ-asymmetric} formally coincides with the zero-temperature Ambegaokar-Baratoff formula \cite{AmbegaokarBaratoff}.

Next, we analyze the quasiparticle contribution Eq.~\eqref{LJqp-expression} by switching to the integration over $\epsilon_{L,R}$, collecting terms before $f_{L,R}$ and taking one of the two energy integrals in the principal value sense, which is allowed since the whole expression \eqref{LJqp-expression} is regular at $\epsilon_L \sim \epsilon_R$. This yields
\be \label{LJ-qp}
	\frac{1}{L_J^{qp}} = - 4 G_N  \int_{\Delta_L}^{\Delta_R} d\epsilon \frac{  \Delta_L \Delta_R}{\sqrt{\epsilon^2 - \Delta_L^2}\sqrt{\Delta_R^2 - \epsilon^2}} f_{\epsilon},
\ee
where we assume $\Delta_L \leq \Delta_R$ without loss of generality. Note that integration in \eqref{LJ-qp} involves quasiparticles with the energy limited by $\delta\Delta$ and/or typical quasipartcile energy $\delta E$ due to the $f_{\epsilon}$ factor. In the following Section, we will evaluate Eq.~\eqref{LJ-qp} assuming energy distribution of quasiparticles with effective temperature $T$ (and non-equlibrium density $x_{qp}$) in the limit $\omega,T,\delta\Delta \ll \Delta$, see Eq.~\eqref{L-qp-evaluation}.

For a symmetric junction, $\Delta_L = \Delta_R = \Delta$, the expressions above are simplified. Kinetic inductance in the absence of quasiparticles, Eq.~\eqref{LJ-asymmetric}, simplifies to
\be \label{LJ-value}
	\frac{1}{L_J} = \pi G_N \Delta 
\ee
while the quasiparticle contribution \eqref{LJ-qp} boils down to \cite{Catelani-qp}
\be \label{LJ-xqpA}
	\frac{1}{L_J^{\text{qp}}} = - \frac{2 x_{qp}^A}{L_J}, \qquad x_{qp}^A = f_{\Delta},
\ee
where $x_{qp}^A$ can be interpreted as the population of the Andreev levels merging with the quasiparticle continuum at $\varphi = 0$. Note that the standard finite-temperature Ambegaokar-Baratoff expression~\cite{AmbegaokarBaratoff} involves a factor $\tanh \Delta / (2 T)$. Its expansion in the limit $T/\Delta \ll 1$ leads to Eq.~\eqref{LJ-xqpA} where $f_\Delta$ is the thermal distribution function. Equation~\eqref{LJ-value} also provides the leading contribution in the regime $\delta\Delta \ll \Delta_{L,R}$. 

In the following Sections, we will study the finite-frequency ac response characterized by the admittance $Y(\omega)$ for a junction with $\delta\Delta \neq 0$. 
Its behavior in the $\omega \rightarrow 0$ limit should reproduce the dc physics described in this section. Indeed, the stationary Josephson effect is manifested as a $1/(i\omega)$ pole in the $Y(\omega)$ with the kinetic inductance dictating its residue, see Eq.~\eqref{Y-three-terms}.

\section{Finite-frequency Admittance of an asymmetric Josephson junction}
\label{sec:finiteomega}

Linear current response of a system to an applied voltage at a finite frequency is characterized by its complex admittance $Y$. According to the Josephson relation, applying voltage at angular frequency $\omega$ can be represented by a periodic modulation of the superconduting phase difference across the junction:
\be \label{phi-perturbation}
	\varphi(t)  = \varphi_0 + \frac{2eV}{\omega} \sin \omega t,
\ee
where $V$ is the voltage amplitude. 

To obtain the linear response of the junction in the leading order we resort to the perturbative expansion in $V$ and average tunneling $\bar{t^2}$. Technically, one can perform a calculation in the Kubo formalism (see Appendix~\ref{sec:Kubo-derivation}) or evaluate the dissipative part first by applying Fermi's golden rule \cite{CatelaniGlazman-qp-SciPost}. Then, the non-dissipative part governing corrections to the kinetic inductance and resonance frequency can be obtained from the Kramers-Kronig formula complemented by matching\cite{LandauLifshitz-StatMech} the purely inductive $1/(i\omega)$ term with the one obtained from the dc physics described in the previous Section. The resulting expression has the following structure:
\be \label{Y-three-terms}
    Y = \left( \frac{1}{i \omega L_J} + \frac{1}{i \omega L_J^{qp}} \right) \cos \varphi_0 + \tilde{Y} \frac{1 + \cos\varphi_0}{2}.
\ee
Here $L_J$ and $L_J^{qp}$ are given by Eqs.~\eqref{LJ-asymmetric} and \eqref{LJ-qp} respectively, and $\tilde{Y}$ is the dynamical quasiparticle contribution evaluated below. The $1/(i\omega)$ part in this formula is adiabatically connected to the dc Josephson effect, while the last term is a regular function of $\omega$ at $\omega\to 0$, provided that $\Delta_L\neq\Delta_R$.

\subsection{Dissipative part of the admittance}
\label{sec:admittance-Re}

The dissipative part of the admittance is most easily obtained~\cite{CatelaniGlazman-qp-SciPost} by equating the dissipation rate $ (V^2/2) \Re Y$
with the one obtained from the Fermi golden rule in the leading order in $V$. 

Note that dissipative processes at $\omega>\Delta_L+\Delta_R$ include the processes of Cooper pair breaking, along with the ones preserving the net number of quasiparticle excitations.
At $\omega < \Delta_L+\Delta_R$ only the latter processes are allowed by the energy conservation. Since the regime of the frequencies that do not exceed $\Delta_{L,R}$ is the most relevant experimentally, we will make this assumption and present here the expressions that do not involve quasiparticle generation by the perturbation Eq.~(\ref{phi-perturbation}). The full result, Eq.~\eqref{ReY-full}, is given in Appendix \ref{sec:Kubo-derivation}. 

To proceed with the analytic calculation, we will additionally take experimentally relevant assumptions
$\omega, \delta E \ll \Delta_{L,R}$ (where $\delta E$ is the typical quasiparticle energy; for a quasi-thermal distribution characterized by an effective temperature $T$, we replace $\delta E$ by $T$). 
A gap difference $\delta\Delta$ prevents low-energy quasiparticles from crossing the junction. That leads to an exponential suppression of dissipation at $\omega<\delta\Delta$ once the temperature of the quasiparticles is low enough, $T\ll\delta\Delta-\omega$. 
The full form of the junction response function is quite cumbersome, see Appendix~\ref{sec:Kubo-derivation}, so in the rest of the main text of this paper we assume the gap difference small,
\begin{equation}
\frac{\delta\Delta}{\Delta}\equiv\frac{\Delta_R-\Delta_L}{\Delta}\ll1,\quad \Delta\equiv\frac{\Delta_L+\Delta_R}{2}
\label{eq:deltaDelta}
\end{equation}
(we set $\Delta_R>\Delta_L$, for definiteness).

This assumption allows us to replace the expressions for $u_n$ and $v_n$ with their value at the gap edge, $1/\sqrt{2}$ and write down the dissipative part of the admittance in the form
\be \label{ReY-through-ReYtilde}
    \Re Y = \Re \tilde{Y} \cdot \frac{1 + \cos \varphi_0}{2},
\ee
where $\Re \tilde{Y}$ is
\be \label{ReY-result}
    \Re \tilde{Y}(\omega) = \frac{e^2}{2\omega} \left[S_{qp}(\omega) - S_{qp}(-\omega)\right].
\ee
The expression for the spectral density $S_{qp}(\omega)$ reads:
\begin{align}
    S_{qp}(\omega) &= \frac{16E_J}{\pi\Delta}\int_{\Delta_{L,R}}^\infty{} d\epsilon_L  d\epsilon_R \frac{\epsilon_L \epsilon_R f_{L} (1-f_{R}) \delta_{\epsilon_{L}-\epsilon_{R} + \omega}
    }{\sqrt{\epsilon_L^2 - \Delta_L^2}\sqrt{\epsilon_R^2 - \Delta_R^2}}
    \nonumber \\
    & \qquad \quad + \text{L $\leftrightarrow$ R},
    \label{S-qp-unexpanded-integrals}
\end{align}
where $E_J = \hbar^2 / (4 e^2 L_J)$ is the Josephson energy expressed through the no-quasiparticle kinetic inductance $L_J$ given by Eq.~\eqref{LJ-value}; $\delta_X \equiv \delta(X)$ denotes the delta-function, and the integration over $\epsilon_{L(R)}$ is performed over all values of $\epsilon_{L(R)} > \Delta_{L(R)}$. 
Note that the the terms containing cross-products $f_L f_R$ cancel from Eq.~\eqref{ReY-result}, but even when evaluating $S_{qp}$ alone, one may neglect the cross-products under the made assumption of low quasiparticle density $x_{qp} \ll 1$.

Since $\delta \Delta \ll \Delta$, we simplify the denominators revealing the singular density of states near the gap edge:
\be \label{dos-singular-approximation}
    \frac{\epsilon_{L(R)}}{\sqrt{\epsilon_{L(R)}^2 - \Delta_{L(R)}^2}} \sim \sqrt{\frac{\Delta}{2\varepsilon_{L,R}}},
\ee
where $\varepsilon_{L(R)} = \epsilon_{L(R)} - \Delta_{L(R)}$. Then, the spectral density reads
\begin{align}
    S_{qp}(\omega) &= \frac{8 E_J}{\pi} \int_0^{\infty} d\varepsilon \frac{f_{\Delta_L + \varepsilon}\theta_{\varepsilon+\omega+\Delta_L -\Delta_R}}{\sqrt{\varepsilon(\varepsilon+\omega + \Delta_L - \Delta_R)}} 
 + \text{L $\leftrightarrow$ R},
    \label{Sqp-through-integrals}
\end{align}
where $\theta_x$ is the Heaviside theta-function. 

As mentioned above, a practically important case is when the quasiparticle distribution function $f_{\epsilon}$ is quasi-thermal, i.e. the one that has an effective temperature $T$ but a non-equilibrium number of quasiparticles $x_{qp}$, which determines some effective chemical potential $\mu_*$. Additionally, it has been discussed that typical values $f_{\epsilon} \ll 1$, so that Fermi-Dirac distribution boils down to the Boltzman form:
\be \label{thermal-f}
f_{\Delta_{L}+\varepsilon_{L}}=f_{0}e^{-\varepsilon_{L}/T}, \  f_{\Delta_{R}+\varepsilon_{R}}=f_{0}e^{-\delta\Delta/T - \varepsilon_{R}/T},
\ee
where $f_0 = f_{\Delta_L}$ is the value of the distribution function at the lower gap edge. 
For the distribution function Eq.~\eqref{thermal-f}, one can analytically take the integrals in Eq.~\eqref{S-qp-unexpanded-integrals} arriving at the following result for positive frequencies ($\omega > 0$), corresponding to the upward transition of the quasiparticle:
\begin{align} \label{S-qp-result}
    S_{qp}(\omega) &  = \frac{8 E_J f_{0}}{\pi} \Big[ e^{-\delta\Delta/T}H\left(\frac{\omega+\delta\Delta}{T}\right) \\ \nonumber
    & + H\left(\frac{|\omega-\delta\Delta|}{T}\right)\left(\theta_{\omega-\delta\Delta}+e^{-(\delta\Delta-\omega)/T}\theta_{\delta\Delta-\omega}\right) \Big] .
\end{align}
For negative frequencies, representing the quasiparticle relaxation,
\be \label{S-qp-minus-omega}
    S_{qp}(-\omega) = S_{qp}(\omega)e^{-\omega/T}.
\ee
Here $H(x) = e^{x/2} K_0(x/2)$, where $K_0$ is the modified Bessel function of the second kind. The function $H(x)$ has the following asymptotic behavior:
\be \label{H-asymptotics}
 H(x) \sim \begin{cases}
 	-\gamma_{G}-\log\frac{x}{4}+\dots,&x\ll1, \\
	\sqrt{\frac{\pi}{x}}\left(1-\frac{1}{4x}+\dots\right),&x\gg1,
 \end{cases}
\ee
where $\gamma_G$ is the Euler–Mascheroni constant. 

Note that the second term of Eq.~\eqref{S-qp-result} exhibits a logarithmic divergence at $\omega = \delta\Delta$ coming from the integral of the product of two density-of-states factors, each diverging as $1/\sqrt{\varepsilon}$.
According to Eq.~\eqref{ReY-result}, $\Re Y$ is obtained by multiplying Eq.~\eqref{S-qp-result} by a factor $e^2(1 - e^{-\omega/T})/2\omega$, which is regular at $\omega\to0$. We present this result graphically in Fig.~\ref{fig:Y-one-asymm}(c), expressed in terms of the average quasiparticle density $x_{qp}$ [Eq.\eqref{f0-via-xqp}], and analyze its asymptotic behavior below.

\subsection{Non-dissipative part of the admittance}
\label{sec:non-dissipative-part}

The non-dissipative (imaginary) part of the dynamical admittance can be obtained from its dissipative (real) part through the Kramers-Kronig relation. However, this method alone cannot capture the purely inductive part, which has a $1/(i\omega)$ frequency dependence~
\cite{LandauLifshitz-StatMech}. This part can be restored by matching it to the $\omega\to 0$ limit, see Eq.~\eqref{Y-three-terms}.
The resulting expression coincides with the brute-force calculation in the Kubo formalism presented in      Appendix~\ref{sec:Kubo-derivation}.

We start by applying the Kramers-Kronig formula to Eqs.~\eqref{ReY-result}--\eqref{S-qp-unexpanded-integrals} and obtain the following contribution to the complex admittance:
\begin{align} \label{Y1-expression}
	\tilde{Y} & = \frac{2G_N}{\pi} \int_{\Delta_{L(R)}}^{\infty} d\epsilon_L d\epsilon_R \frac{
    \epsilon_L \epsilon_R 
    }{
    \sqrt{\epsilon_L^2 - \Delta_L^2}\sqrt{\epsilon_R^2 - \Delta_R^2}
    } \frac{f_R - f_L}{\epsilon_R - \epsilon_L}
    \nonumber \\ 
    & \quad \qquad \times 
    \frac{i}{\omega-\epsilon_{L}+\epsilon_{R}+i 0^{+}} + \text{L $\leftrightarrow R$}.
\end{align}
Remember, however, that Eq.~\eqref{S-qp-unexpanded-integrals} was written at $\omega < \Delta$ and the terms corresponding to the cross-gap transitions were omitted. Therefore, Eq.~\eqref{Y1-expression} also does not include the corresponding parts of the response. That is not a problem, since the latter carry an additional $\omega / \Delta$ factor, so Eq.~\eqref{Y1-expression} is sufficient in the limit $\omega \ll \Delta$. For completeness, the full expression at arbitrary $\omega / \Delta$ is given in \eqref{Y-full-appendix}.

In the previous section, we argued that at $\omega,T\ll \delta\Delta$ and fixed density $x_{qp}$, dissipation associated with the quasiparticle tunneling is suppressed exponentially in parameter $\delta\Delta/T$. Correction to the non-dissipative part of response is associated with the virtual transitions, and therefore under the same conditions is less sensitive to the quasiparticle energy distribution function. 
Below we will focus on the limit $\omega,T,\delta\Delta\ll\Delta_{L,R}$ and arbitrary ratios between $\omega$, $T$, and $\delta\Delta$, relegating the most general case to Appendix~\ref{sec:Kubo-derivation}, see Eq.~\eqref{Y-full-appendix}.
As in the previous section, the condition $\delta\Delta\ll \Delta$ justifies replacing $u_n$ and $v_n$ amplitudes with $1/\sqrt{2}$ [which has been done in obtaining Eq.~\eqref{Y1-expression}] and approximation \eqref{dos-singular-approximation} for the density-of-states prefactors. We switch again to the integration over $\varepsilon_{L(R)} = \epsilon_{L(R)} - \Delta_{L(R)}$, apply a simple partial fraction decomposition, and  integrate each term over one of the two variables $\varepsilon_{L}$ or $\varepsilon_{R}$---whichever one does not enter as the argument of the distribution function. After some algebra,
we cast the result for the imaginary part in the following form:
\begin{widetext}
${}$
\begin{align}
& \frac{\Im Y} {(\omega L_J)^{-1}} 
    = - \cos \varphi_0 + \frac{1 + \cos \varphi_0}{2} \left( 
    \int_{0}^{|\omega-\delta\Delta|} \frac{d\varepsilon}{\sqrt{\varepsilon}} \frac{ f_{\Delta_{R}+\varepsilon} \theta_{\omega-\delta\Delta} + f_{\Delta_{L}+\varepsilon} \theta_{\delta\Delta-\omega}}{\pi\sqrt{|\omega-\delta\Delta|-\varepsilon}}  
    \nonumber + \int_{0}^{\omega+\delta\Delta} \frac{d\varepsilon}{\sqrt{\varepsilon}} \frac{f_{\Delta_{L}+\varepsilon}}{\pi \sqrt{\omega+\delta\Delta-\varepsilon}} \right)
    \\
    & \qquad \qquad \qquad \ \ - \frac{1 - \cos\varphi_0}{2}
    \int_{0}^{\delta\Delta} \frac{d\varepsilon}{\sqrt{\varepsilon}} \frac{2f_{\Delta_{L}+\varepsilon}}{\pi \sqrt{\delta\Delta-\varepsilon}}.
\label{Im-Y-general}
\end{align}
For the quasi-thermal distribution function \eqref{thermal-f}, the integrals above can be evaluated analytically yielding
\begin{align} 
	& \Im Y = 
    - \frac{\cos \varphi_0}{\omega L_J} + \frac{f_{0}}{\omega L_{J}} 
\left\lbrace \frac{1 + \cos \varphi_0}{2} \left[ h\left(\frac{\omega+\delta\Delta}{T}\right) 
   + (\theta_{\delta\Delta-\omega} + \theta_{\omega-\delta\Delta}e^{-\delta\Delta/T})h\left(\frac{|\omega-\delta\Delta|}{T}\right) \right] \right.
 \nonumber \\
   & \qquad \qquad \qquad \qquad \qquad \ \ -
   \left. \frac{1 - \cos \varphi_0}{2} 2h\left(\frac{\delta\Delta}{T}\right)\right\rbrace.
    \label{ImY-result}
\end{align}
\end{widetext}
Here $h(x)=e^{-x/2}I_{0}(x/2)$, where $I_0$ is the modified Bessel function of the first kind. The function $h(x)$ has the following asymptotes:
\be \label{h-asymptotics}
	h(x) \sim \begin{cases}
1-\frac{x}{2}+\dots, & x\ll1,\\
\frac{1}{\sqrt{\pi x}} \left(1+\frac{1}{4x}+\dots\right), & x\gg1.
\end{cases}
\ee

Equations \eqref{Im-Y-general} and \eqref{ImY-result} represent the full non-dissipative part of the junction admittance at a phase bias $\varphi_0$. Note that $\Im Y(\omega)\propto\cos \varphi_0 / (i\omega)$ in the limit $\omega\to 0$, in agreement with the first term in Eq.~\eqref{Y-three-terms}, where $L_J^{qp}$ is given by
\be \label{L-qp-evaluation}
    - \frac{1}{L_J^{qp}} = \frac{1}{\pi L_J} \int_{0}^{\delta\Delta} \frac{d\varepsilon}{\sqrt{\varepsilon}} \frac{2f_{\Delta_{L}+\varepsilon}}{\sqrt{\delta\Delta-\varepsilon}} = \frac{2 f_0}{L_J} h\left(\frac{\delta\Delta}{T}\right).
\ee
The integral in Eq.~\eqref{L-qp-evaluation} is restricted to the energy interval between the gaps in the leads, $\Delta_L$ and $\Delta_R$. In the limit $\delta\Delta \rightarrow 0$, it produces the value of the distribution function at the gap edge, $f_{\Delta} = x_{qp}^A$ as was preempted in Eq.~\eqref{LJ-xqpA}, which is valid even for a non-thermal $f_\epsilon$. Note  also that at $\varphi_0 = 0$, the last term in Eqs.~\eqref{Im-Y-general} and \eqref{ImY-result} vanish, which was noted in Ref.~\onlinecite{Catelani-qp} as the $x_{qp}^A$-term cancellation.
The two scales for the temperature dependence of the remaining correction to the admittance are $\delta \Delta + \omega$ and $|\delta \Delta - \omega|$.  

We show the frequency dependence of $\Re Y$ and $\Im Y$ in Fig.~\ref{fig:Y-one-asymm}(c, d) for  zero phase bias, $\varphi_0 = 0$, expressed in terms of the average quasiparticle density $x_{qp}$ [Eq.~\eqref{f0-via-xqp}]. Since $Y(\omega)$ is an analytic function of $\omega$, the logarithmic singularity of its real part at $\omega = \delta \Delta$ is accompanied by a jump in the imaginary part. The jump is pronounced the most at $T\ll\delta\Delta$, as seen from the second term within the square brackets in Eq.~\eqref{ImY-result}. In the following Section we analyze its impact on the frequency shift of a transmon.

\begin{figure} \centering\includegraphics[width=0.48\textwidth]{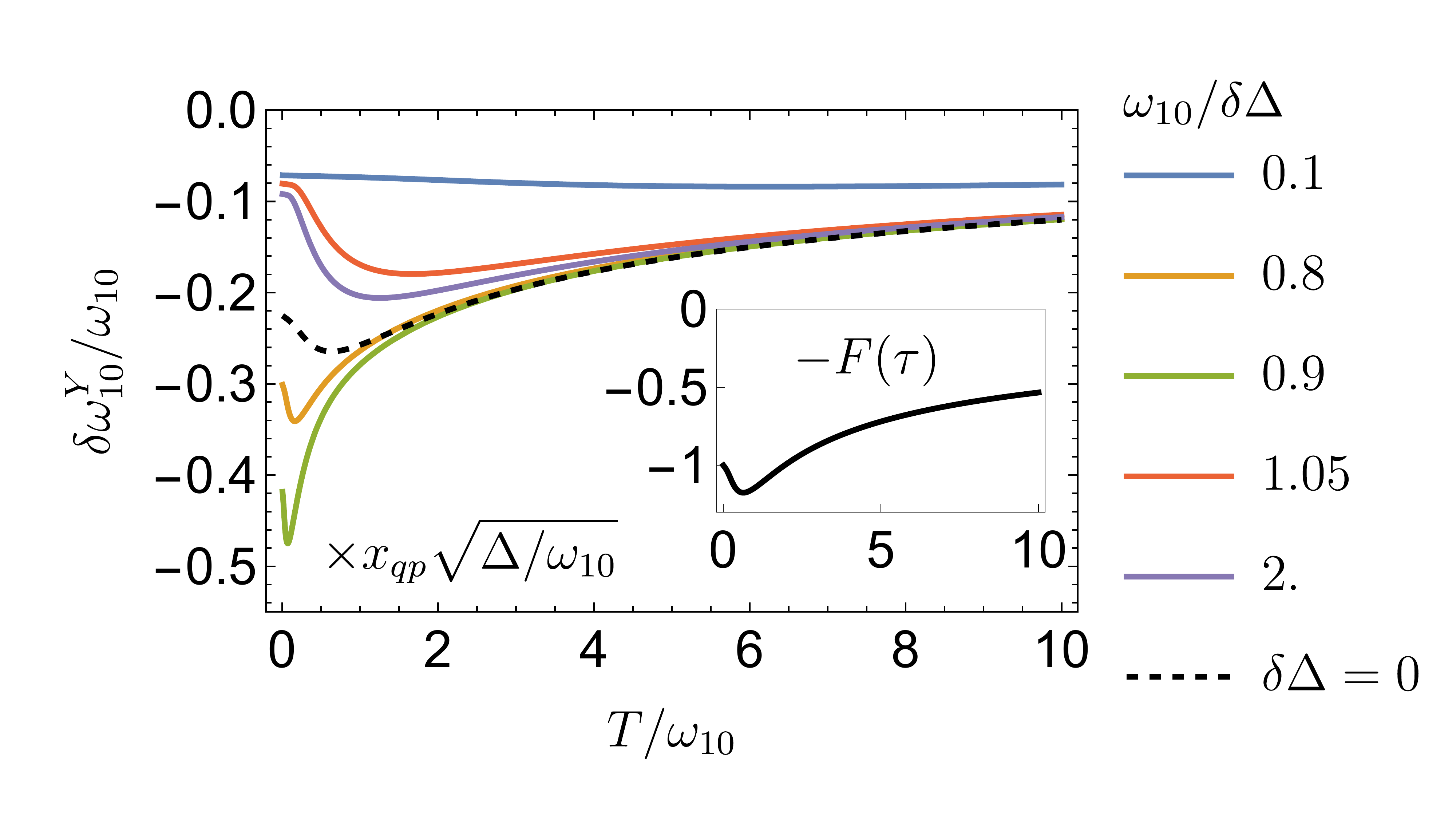}
    \caption{
    Temperature dependence of the relative frequency shift of the split-transmon at $\Phi = 0$ due to quasiparticles, $\delta \omega_{10}^Y / \omega_{10}$, as given by Eq.~\eqref{delta-omega-Y-result}; the practically relevant limit $\mathcal{V}_L \gg \mathcal{V}_R$ was taken so that $\zeta(T) = 1$. The plot uses units of $x_{qp} \sqrt{\Delta / \omega_{10}}$ and shows results for various values of $\omega_{10} / \delta\Delta$, compared to the symmetric junction case, $\delta\Delta = 0$ (black dashed line). All curves exhibit non-monotonic behavior, but vary on a different temperature scale: $T \sim \omega_{10}$ when $\delta\Delta < \omega_{10}$ (red, purple), $T \sim \delta\Delta$ when $\omega_{10} \ll \delta\Delta$, with suppressed amplitude (blue), and $T \sim T_* = \delta\Delta - \omega_{10}$ when $0 < \delta \Delta - \omega_{10} < \delta\Delta$, resonantly enhanced amplitude (green, orange). In the latter case, the rescaled curves collapse on a universal form $-(x_{qp} / \pi) \sqrt{\Delta / (2 T_*)} F(T / T_*)$, see Eq.~\eqref{low-T-asymptotics}; the function $-F(\tau)$ is shown in the inset. Note the non-monotonic progression of the amplitude with $\omega_{10}/\Delta$, cf. the $\Im Y(\omega)$ frequency dependence shown in Fig.~\ref{fig:Y-one-asymm}.\label{fig:T-depedence}
    }
\end{figure}

\section{
Shift of a split-transmon frequency induced by quasiparticles
}
\label{sec:frequency-shift}

A split-transmon consists of a loop of two Josephson junctions (with Josephson energies $E_{J1}, E_{J2}\gg E_C$) pierced by magnetic flux $\Phi$. Its Hamiltonian reads
\be \label{H-split-transmon}
    H_{\phi} = 4E_C \hat{N}^2 - E_{J1} \cos \left( \hat{\phi} - 2\pi\frac{\Phi}{\Phi_0} \right) - E_{J2} \cos \hat\phi,
\ee
where $\hat{N} = -i\partial/\partial \phi - n_g + (P-1) / 4$ is the operator of the Cooper pair number\cite{Serniak-DirectDispersiveMonitoring}, which involves the dimensionless offset charge $n_g$ and the parity $P = \pm 1$. Next, 
$\Phi_0 = \pi \hbar / e$ is the superconducting flux quantum, and the charging energy $E_C = e^2 / C_\Sigma$, where $C_\Sigma$ is the total capacitance of the circuit. 
The kinetic inductance of each junction, $L_{J,j}$, is expressed through its Josephson energy as $L_{J,j} = 1 / (4 e^2  E_{J,j})$. Note that away from the special point $\Phi = \Phi_0/2$, or at any flux if $|E_{J1}-E_{J2}|\gg E_C$, stray charges produce only exponentially small corrections, so we ignore $n_g$ from now on. 

The limit of large Josephson coupling justifies expansion in the vicinity of the classical ground state $\hat\phi = \pi \Phi/\Phi_0 + \vartheta$, where $\vartheta$ is defined by the relation
\be \label{vartheta-expression}
    \tan \vartheta = \frac{E_{J1} - E_{J2}}{E_{J1} + E_{J2}} \tan\frac{\pi \Phi}{\Phi_0}.
\ee
Substituting $\hat\phi = \pi \Phi/\Phi_0 + \vartheta + \hat\varphi$, we get
\be \label{H-split-transmon-shifted}
    H_{\varphi} = 4E_C \hat{N}^2 - \sum_{j=1,2} E_{J_j} \cos \left( \hat{\varphi} - (-1)^j  \frac{\pi \Phi}{\Phi_0} - \vartheta \right).
\ee
The spectrum of this system consists of a sequence (labeled $0,1,\dots$) of well-separated and nearly degenerate ($e/o$ for even/odd, $P = \pm1$) levels of opposite parity. Moreover, under the assumed  conditions, the zero-point fluctuations of phase across each junction, $\varphi_{\rm zpf,j}\sim (E_C/E_{Jj})^{1/4}$, are weak, so that the anharmonicity effect on the $|0\rangle-|1\rangle$ transition frequency $\omega_{10}$ is small, allowing us here to use
\be
    \omega_{10}(\Phi) = \sqrt{8 E_C E_J (\Phi)},
\ee
with
\be
    E_J^2 = (E_{J1} + E_{J2})^2 \cos^2 \frac{\pi \Phi}{\Phi_0} + (E_{J1} - E_{J2})^2 \sin^2 \frac{\pi \Phi}{\Phi_0} .
\ee

\subsection{Quasiparticle corrections}

The presence of quasiparticles affects $\omega_{10}$. The respective frequency shift can be related to the quasiparticle contributions to the junctions admittances.
The total admittance of the Josephson junctions in the split-transmon is the sum of the individual contributions of the two junctions:
\be
    Y(\omega) = Y_1(\omega) + Y_2(\omega),
\ee
which, together with the total capacitance $C_{\Sigma}$, determines the transmon frequency. Quasiparticles affect the junction admittances, $Y_{0,j}(\omega)\to Y_{0,j}(\omega)+\delta Y_{0,j}(\omega)$, leading to the split transmon frequency shift,
\be \label{delta-omega-admittance}
    \delta \omega_{10}^Y = - \frac{\omega_{10}^2(0)}{2} \frac{\sum_{j=1,2}\Im \delta Y_j (\omega_{10}, \varphi_{0,j})}{\sum_{j=1,2} L_{J,j}^{-1} }.
\ee
The phase biases in this equation are given by
\be
    \varphi_{0,j} = \vartheta \pm \frac{\pi \Phi}{\Phi_0},
\ee
with $\vartheta$ defined in Eq.~\eqref{vartheta-expression}.

In practical cases, where $x_{qp} \ll 1$, the resulting corrections to $\omega_{10}$ are small and, to linear order, can be separated into two parts:
\be \label{delta-omega-two-parts}
    \delta\omega_{10} = \delta\omega_{10}^\Delta (x_{qp})  +\delta\omega_{10}^Y(x_{qp}, T). 
\ee
The first part comes from the dependence of the gap on $x_{qp}$, see Eq.~(\ref{gap-suppression}); it is insensitive to temperature as long as it remains far below the transition temperature value, $T\ll T_c$, when $x_{qp}$ is wholly determined by the resident quasiparticles. The second part comes from the variation of the impedance with the density and temperature of quasiparticles~\cite{Catelani-qp} at a fixed gap value; the characteristic scales for $T$ are controlled by $\omega_{10}$, $\delta\Delta$, and $|\delta\Delta - \omega_{10}|$, all much smaller than $T_c$. The temperature dependence of the latter part is studied in Section~\ref{sec:temperature-dependence}. The effect of quasiparticle-induced frequency shift $\delta\omega_{10}^\Delta$ through the reduction of the superconducting gap is discussed in Section~\ref{sec:omega-delta-and-xqpT}.

In the following, we will assume that both junctions in the split-transmon have the same gap difference in the leads $\delta\Delta$. At the same time, we will allow for the unequal Josephson energies $E_{J,j}$ due to different tunneling conductances. 

\subsection{Temperature dependence $\delta\omega_{10}^Y(x_{qp}, T)$ 
}
\label{sec:temperature-dependence}

The frequency shift $\delta \omega_{10}^Y$ is negative at all parameter values but has a non-trivial temperature dependence, which we discuss below.

Before diving into details, we summarize the main features that $\delta\Delta\neq 0$ introduces into the temperature dependence, see also Fig.~\ref{fig:T-depedence}. In the absence of gap difference, $\delta\Delta=0$, the transmon frequency $\omega_{10}$ is the only relevant scale for the variation of  $\delta\omega_{10}^Y(x_{qp}, T)$ with temperature. On this scale, the relative frequency shift $\delta\omega_{10}^Y(x_{qp}, T)/\omega_{10}$ starts with a low-temperature negative value, $\delta\omega_{10}^Y/\omega_{10} \approx -x_{qp} \sqrt{\Delta/\omega_{10}}$, at $T=0$, becomes about $10\%$ more negative at $T \sim \omega_{10}$, and then slowly increases, approaching zero in the high-temperature limit, $\delta\omega_{10}^Y/\omega_{10} \approx -x_{qp} \sqrt{\Delta/T}$, at $T \gg \omega_{10}$. In contrast, a finite gap difference $\delta\Delta$ may affect the typical values of $\delta\omega_{10}^Y$ and introduce an additional scale for its  temperature dependence. 

We will show below that in the case of a large gap difference, $\delta\Delta\gg\omega_{10}$, the dependence of $\delta\omega_{10}^Y$ on $T$ retains its shape, but with a suppressed overall variation of frequency, $\delta\omega_{10}^Y(x_{qp}, 0)\approx -x_{qp}\sqrt{\Delta/\delta\Delta}$; the temperature scale over which $\delta\omega_{10}^Y$ varies is extended from $\omega_{10}$ to $\delta\Delta$ (beyond this scale, the effect of $\delta\Delta\neq 0$ is insignificant).

Tuning $\omega_{10}$ close to $\delta\Delta$ from below introduces a new temperature scale, $T_\star=\delta\Delta-\omega_{10}>0$. At $T_\star\ll\delta\Delta$, it is this new scale that determines the most significant variation of $\delta\omega_{10}^Y/\omega_{10}$. Compared to the case of $\delta\Delta=0$, the frequency shift is parametrically stronger, $\delta\omega_{10}^Y/\omega_{10}\sim -x_{qp}\sqrt{\Delta/T_\star}$ and narrower, demonstrating a minimum at $T\sim T_\star$. At higher temperatures, the frequency shift slowly reaches zero, roughly the same way as in the case $\delta\Delta=0$.

In the derivation of $\delta\omega_{10}^Y$ at low temperatures, $T\ll T_c$, we assume that the average density of quasiparticles
\be \label{xqp-through-nqp}
    x_{qp} = \frac{N_{qp}^L + N_{qp}^R}{2\nu_0 \Delta (\mathcal{V}_L + \mathcal{V}_R)}
\ee
remains fixed~\footnote{We expect this condition to hold for a source of nonequilibrium quasiparticles that is spatially uniform on the scale of the transmon size, such as phonons generated by a cosmic-ray impact\cite{qp-poisoning-Plourde} or on-chip phonon emitters \cite{PhononDowncoversion-Plourde-Nature}.}
while their chemical potential varies with $T$; here $\mathcal{V}_{L(R)}$ are the volumes of the material with lower (higher) gap. The number of quasiparticles in each of the two materials 
can be calculated in the approximation Eq.~\eqref{dos-singular-approximation}  as
\be \label{nqp-integral}
    N_{qp}^{L(R)} = 4 \mathcal{V}_{L(R)} \int_{0}^{\infty}d\varepsilon_{L(R)}\sqrt{\frac{\Delta}{2\varepsilon_{L(R)}}}f_{\Delta_{L(R)} + \varepsilon_{L(R)}},
\ee
where the factor $4$ accounts for spin and particle-hole degeneracies. Substituting \eqref{thermal-f} and doing some algebra, we obtain
\be \label{f0-via-xqp}
	f_0 = x_{qp} \sqrt{\frac{\Delta}{2\pi T}} \zeta(T), \quad \zeta(T) = \frac{\mathcal{V}_L + \mathcal{V}_R}{\mathcal{V}_L + \mathcal{V}_R e^{-\delta \Delta / T}}.
\ee
Substitution of Eq.~\eqref{f0-via-xqp} into Eq.~\eqref{ImY-result} produces an expression  for the imaginary part of the admittance with an explicit temperature dependence. 
After that, we use Eq.~\eqref{delta-omega-admittance} to find the frequency shift $\delta \omega_{10}^Y$  in the form:
\begin{widetext}
\begin{align} \label{delta-omega-Y-result}
    \delta \omega_{10}^Y (\Phi) = 
    - \frac{x_{qp} \zeta(T) \omega_{10}(\Phi)}{2} \sqrt{\frac{\Delta}{2\pi T}} 
  \left\lbrace 
\frac{\omega_{10}(0)^2 + \omega_{10}(\Phi)^2}{2\omega_{10}(\Phi)^2} (\theta_{\delta\Delta-\omega_{10}(\Phi)} + \theta_{\omega_{10}(\Phi)-\delta\Delta}e^{-\delta\Delta/T})h\left(\frac{|\omega_{10}(\Phi)-\delta\Delta|}{T}\right)\right.
\\ \nonumber 
\left. - \frac{\omega_{10}(0)^2 - \omega_{10}(\Phi)^2}{\omega_{10}(\Phi )^2} h\left(\frac{\delta\Delta}{T}\right) 
     + \frac{\omega_{10}(0)^2 + \omega_{10}(\Phi)^2}{2\omega_{10}(\Phi)^2}
h\left(\frac{\omega_{10}(\Phi)+\delta\Delta}{T}\right)\right\rbrace,
\end{align}
\end{widetext}
where $\omega_{10}(0)$ is the transmon frequency at zero flux, function $h(x)$ is defined above Eq.~\eqref{h-asymptotics}, and $\zeta(T)$ is the volume-dependent factor from Eq.~\eqref{f0-via-xqp}.

We now proceed to analyze the special limiting cases.
First, consider the case where the gap difference is much larger than the frequency, $\omega_{10} / \delta\Delta \ll 1$. In this case, Eq.~(\ref{delta-omega-Y-result}) simplifies to:
\begin{align} \label{wmmdd-asymptotics}
    \delta \omega_{10}^Y & =
    -x_{qp}\zeta(T) \omega_{10}(\Phi) \sqrt{\frac{\Delta}{2\pi T}} h\left(\frac{\delta\Delta}{T} \right),
\end{align} 
This contribution has the same form as the inductance renormalization \eqref{L-qp-evaluation} and is accompanied by a suppressed dissipation. 
(These statements are valid even if the distribution function $f_{\epsilon}$ deviates from the quasithermal one provided that the typical quasiparticle energy $\delta E \ll \Delta$, irrespective of the value of $\delta E/\delta\Delta$.)
The variation of frequency with temperature occurs in a broader temperature range, $T\sim\delta\Delta$, than for a symmetric junction ($T\sim\omega$). At $T \ll \delta\Delta$, the typical values of $\delta \omega_{10}^Y/\omega_{10}$ are on the order of $x_{qp}\sqrt{\Delta/\delta\Delta}$, smaller than in a symmetric junction by a factor of $\sqrt{\omega/\delta\Delta}$. At high temperatures, $T \gg \delta\Delta,\omega_{10}$, the asymptotic behavior
\begin{align} \label{high-T-asymptotics} 
    & \delta \omega_{10}^Y = - x_{qp} \omega_{10}(\Phi) \sqrt{\frac{\Delta}{2 \pi T}}
\end{align}
coincides with that of a symmetric junction \cite{Catelani-qp} ($\delta\Delta = 0$). 

Next we consider the case of comparable $\omega_{10}$ and $\delta\Delta$. The temperature dependence of  $\delta\omega^Y_{01}$ in this case is visualized in Fig.~\ref{fig:T-depedence}. The most interesting features appear when $\delta\Delta > \omega_{10}$ and $(\delta \Delta - \omega_{10})/\delta \Delta \ll 1$ introducing a new temperature scale $T_* = \delta\Delta - \omega_{10}$, distinct from the other scales in the system. Assuming additionally $T \ll \delta\Delta$, we may simplify Eq.~(\ref{delta-omega-Y-result}) to
\begin{align} \label{low-T-asymptotics}
     & \delta \omega_{10}^Y =  
    \\ \nonumber
    & \ -\frac{x_{qp} \zeta(T) \omega_{10}(\Phi)}{4\pi\sqrt{2}} 
    \Bigg(\sqrt{\frac{\Delta}{T_*}} F\left(\frac{T}{T_*}\right) \frac{\omega_{10}(0)^2 + \omega_{10}(\Phi)^2}{\omega_{10}(\Phi)^2}
    \\ \nonumber
    & \qquad \qquad \qquad \qquad+ \sqrt{\frac{\Delta}{\omega_{10} + \delta \Delta}} \frac{\omega_{10}(0)^2 + \omega_{10}(\Phi)^2}{\omega_{10}(\Phi)^2} 
    \\ \nonumber
    & \qquad \qquad \qquad \qquad - 2\sqrt{\frac{\Delta} {\delta\Delta}} \frac{\omega_{10}(0)^2 - \omega_{10}(\Phi)^2}{\omega_{10}(\Phi )^2}
    \Bigg),
\end{align}
where $F(\tau) = \sqrt{\pi/\tau} h(1/\tau) = \sqrt{\pi/\tau} e^{-1/(2\tau)} I_0(1/(2\tau))$. 
Here, the dominant contribution comes from the first term, which represents the resonant behavior near the step-like singularity of $\Im Y(\omega_{10})$ at $\omega_{10}=\delta\Delta$ discussed at the end of the previous Section; the other terms are regular and small.
The function $F(\tau)$ is finite at $F(0)=1$, increases by $\sim0.18$ reaching its maximum at $\tau_* \approx 0.63$, and decays as $\sqrt{\pi/\tau}$ at $\tau\gg1$, see the inset in Fig.~\eqref{fig:T-depedence}.
As a result, the relative frequency shift [see the yellow and green lines in Fig.~\ref{fig:T-depedence}] exhibits a dip with a width on the order of $T_*$ and amplitude $x_{qp}\sqrt{\Delta/T_*}$. This dip, which occurs at $T\sim T_*$, is narrower and deeper compared to that in a symmetric junction\cite{Catelani-qp}, where the width is on the order of $\omega$ and the amplitude is $x_{qp}\sqrt{\Delta/\omega}$. 
Following the dip, the frequency shows a long tail, increasing steadily over a broad temperature range, $\delta\Delta - \omega_{10} \lesssim T \ll \delta\Delta, \omega_{10}$. This big increase with temperature is anomalous and constitutes one of the main findings of our work. As the temperature approaches $\delta\Delta$, the resonant term in \eqref{low-T-asymptotics} diminishes and becomes comparable to the non-resonant terms, all of which become small. Finally, at $T \gg \omega_{10}, \delta\Delta$, the frequency shift is further suppressed, and the expansion of the full expression \eqref{ImY-result} becomes independent of the $\omega/\delta\Delta$ ratio, yielding the same
result as \eqref{high-T-asymptotics}.

The resonant enhancement of the frequency shift $\delta\omega^Y_{01}$ discussed above is absent at $\omega_{10} > \delta\Delta$. In that frequency domain, the temperature dependence of $\delta\omega^Y_{01}$ is similar to the $\delta\Delta=0$ case, see the black dashed line in Fig.~\ref{fig:T-depedence}. At the critical value, $\omega_{10} = \delta\Delta$, the frequency changes abruptly by
\begin{align} \label{frequency-jump}
    & \left. \delta \omega_{10}^Y \right|_{\omega_{10} = \delta \Delta + 0} - \left. \delta \omega_{10}^Y \right|_{\omega_{10} = \delta \Delta - 0} = \\ \nonumber
    & \ \ =
    \frac{
    x_{qp} \zeta(T)}{4} \sqrt{\frac{\Delta}{2\pi T}} \left(1 - e^{-\delta\Delta/ T}\right) \frac{\omega_{10}(0)^2 + \omega_{10}(\Phi)^2}{\omega_{10}(\Phi)}.
\end{align}
This abrupt change comes from the jump in the imaginary part of admittance, see Fig.~\ref{fig:Y-one-asymm}(d) and the discussion at the end of Section~\ref{sec:non-dissipative-part}. 
\begin{figure}
    \centering \includegraphics[width=0.9\linewidth]{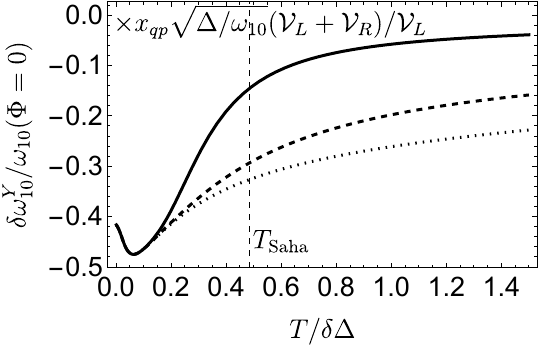}
    \caption{ \label{fig:T-Saha}
    Frequency shift of a split-transmon at $\Phi = 0$ (or a single-junction transmon) at different ratios ${\cal V}_R/{\cal V}_L$. 
    Note that the collapse of the curves at temperatures $T\lesssim T_*$ in the chosen units means that the frequency shift scales as $(\mathcal{V}_L + \mathcal{V}_R) / \mathcal{V}_L$. The shift is strongly enhanced for a small relative volume of the low-gap part of the device. The solid curve is plotted for $\mathcal{V}_L/\mathcal{V}_R= 0.1$. The small ratio $\mathcal{V}_L/\mathcal{V}_R$ gives rise to a new temperature scale, $T_{\rm Saha}$ of Eq.~(\ref{T-Saha}). Two other curves are plotted for $\mathcal{V}_L = \mathcal{V}_R$ (dashed line) and $\mathcal{V}_L / \mathcal{V}_R = 10$ (dotted line). In all the plots, $\omega_{10} / \delta \Delta = 0.9$.}
\end{figure}

We conclude the analysis of the $\omega_{10}^Y(T)$ dependence by considering the case ${\cal V}_L\ll {\cal V}_R$. At a fixed quasiparticle density $x_{qp}$ (averaged over the entire volume ${\cal V}_R+{\cal V}_L$) and low temperature, most of the quasiparticles will accumulate in the lower-gap lead. That strongly enhances their population in the lead, $x_{qp}^{\rm eff}=({\cal V}_R/{\cal V}_L)x_{qp}$, and results in a stronger frequency shift. The non-monotonic part of the $\omega_{10}^Y(T)$ function is enhanced as long as the ``ionization temperature''~\cite{Saha-1, Saha-2},
\be \label{T-Saha}
    T_{\text{Saha}} = \delta\Delta / \log (\mathcal{V}_R / \mathcal{V}_L),
\ee
is high enough, $T_{\text{Saha}}\gtrsim T_*$, see Fig.~\ref{fig:T-Saha}. While such arrangement is unhelpful for the qubit performance, making the frequency shift stronger may be useful for investigating the energy distribution of quasiparticles. 

\subsection{Frequency shift $\delta\omega_{10}^\Delta$ due to the gap suppression by quasiparticles}
\label{sec:omega-delta-and-xqpT}

Here we analyze the first term in \eqref{delta-omega-two-parts}, i.e. the frequency shift $\delta\omega_{10}^\Delta$ caused by the superconducting gap reduction \eqref{gap-suppression} in the presence of quasiparticles. In the linear order in $x_{qp}$, it is obtained by substituting the reduced value of $\Delta$ given by Eq.~\eqref{gap-suppression} into expression \eqref{LJ-value} for $L_{J,j}$ and further into Eq.~\eqref{delta-omega-admittance}. Extracting the part linear in $x_{qp}$ and substituting it in Eq.~\eqref{delta-omega-admittance}, we get after some algebra
\be
    \delta\omega_{10}^\Delta = -\frac{x_{qp, L} + x_{qp, R}}{4}  \omega_{10},
\ee
where $x_{qp,L(R)}$ are individual quasiparticle densities in the left and right lead, see Eq.~\eqref{xqp-definition}. Using \eqref{f0-via-xqp} to express them through $x_{qp}$, we get
\be \label{delta-omega-delta} 
\delta\omega_{10}^\Delta = -\frac{1 + e^{-\delta\Delta / T}}{4} \frac{\mathcal{V}_L + \mathcal{V}_R}{\mathcal{V}_L + \mathcal{V}_R e^{-\delta \Delta / T}} x_{qp} \omega_{10}.
\ee
Note that in the practically relevant regime $T \ll \delta \Delta$ and $\mathcal{V}_L \gg \mathcal{V}_R$, this result reduces to 
\be
    \delta\omega_{10}^\Delta = -\frac{x_{qp}}{4} \omega_{10},
\ee
which agrees with Ref.~\onlinecite{Vlad-new-ErrorBursts}.
Given the working assumptions stated above, $\omega_{10},T,\delta\Delta \ll 1$, the contribution \eqref{delta-omega-delta} is smaller than the admittance-related contribution Eq.~\eqref{delta-omega-Y-result} as the latter contains large factors $\sqrt{\Delta/(\omega \pm \delta\Delta)}$ and $\sqrt{\Delta / T}$, see  Eqs.~\eqref{low-T-asymptotics} and \eqref{high-T-asymptotics}.

As mentioned in the beginning of this Section, at $T \ll T_c$, the resident quasiparticle density, and therefore the shift Eq.~\eqref{delta-omega-delta} is independent of temperature, as in that temperature range $x_{qp}$ is determined by the balance between the quasiparticle generation by external sources and their recombination.
At higher temperatures, the presence of thermal quasiparticles may become dominant. Using Eqs.~\eqref{xqp-through-nqp} and \eqref{nqp-integral} and equilibrium distribution functions, we may introduce
\be \label{xqpT} 
    x_{qp}^T = \sqrt{\frac{2 \pi T}{\Delta}} \frac{\mathcal{V}_L e^{-\Delta_L / T} + \mathcal{V}_R e^{-\Delta_R / T}}{\mathcal{V}_L + \mathcal{V}_R}.
\ee
In the simplest model, they contribute additively so that the total quasiparticle density is
\be \label{xqp-total}
    x_{qp} = x_{qp}^{\text{res}} + x_{qp}^T,
\ee
where $x_{qp}^{\text{res}}$ represents the density of resident quasiparticles. The full frequency shift due to quasiparticles  is given by Eq.~\eqref{delta-omega-two-parts}, with the value of $x_{qp}$ substituted from Eq.~\eqref{xqp-total} and assumed to be small, $x_{qp} \ll 1$.

We conclude this Section with a brief discussion of the effect of anharmonicity. With respect to $\delta\omega_{10}$, its effect is threefold. First, the quasiparticle-induced frequency shift $\delta\omega_{10}$ should be measured with respect to the qubit frequency modified  by the anharmonicity. Second, the modified value of $\omega_{10}$ should be also used in the definition of $T_*=\delta\Delta -\omega_{10}$. Lastly, anharmonicity slightly reduces the overall amplitude of $\delta\omega_{10}$; the relative reduction is~\cite{Catelani-qp} $\sim (E_C/E_J)^{1/2}$.

\section{Transition rates}
\label{sec:transition-rates}

Gap-engineered Josephson junctions with $\delta\Delta \neq 0$ have been proposed as a way to suppress qubit decoherence caused by quasiparticles. In this section, we examine the transition rate at arbitrary ratios $\omega_{10} / \delta\Delta$ and  $T / \omega_{10}$, going beyond the typical split-transmon parameters~\cite{dicarlo_demonstration_2009, Vlad-new-ErrorBursts}, $T_*\sim\delta\Delta$, and operational regime, $T \ll T_*$.

\subsection{Energy relaxation rates}
\label{sec:energyrelaxation}

\begin{figure}[h]
    \vspace{0.25cm}
    \centering \includegraphics[width=0.95\linewidth]{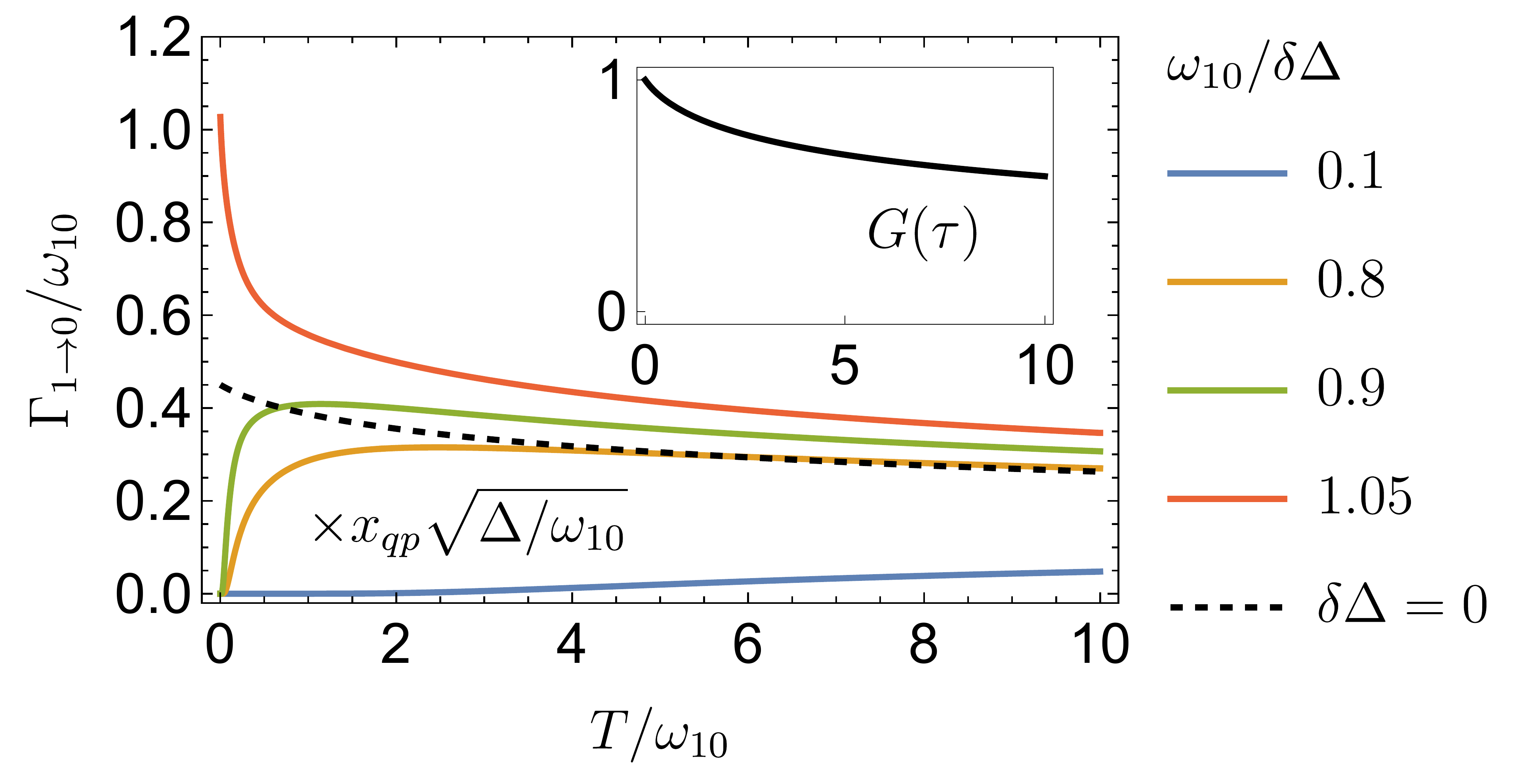} \caption{\label{fig:Gamma-10}  Temperature dependence of the transition rate $\Gamma_{1\to 0}$ of the split-transmon at $\Phi = 0$ due to quasiparticles, as given by Eq.~\eqref{Gamma-downward}; the practically relevant limit $\mathcal{V}_L \gg \mathcal{V}_R$ was taken, so that $\zeta(T) = 1$.
    The plot uses units of $x_{qp} \sqrt{\Delta / \omega_{10}}$ and shows results for various values of $\omega_{10} / \delta\Delta$, which are compared to the symmetric junction case ($\delta\Delta = 0$, black dashed line).
    At $\omega_{10} \ll \delta\Delta$ the dissipation is exponentially suppressed at small temperatures and varies on the scale $\delta\Delta$.
    Near the resonance, $|\delta\Delta - \omega_{10}| \ll \delta\Delta$, the rate is logarithmically enhanced due to the singularity in the density of states. 
    $\Gamma_{1\to 0}$ varies on the scale $T \sim |\delta\Delta - \omega_{10}|$, and is enhanced by the prefactor $(x_{qp}/\pi)\sqrt{\Delta/(2\pi |\delta\Delta - \omega_{10}|)}$, see Eq.~\eqref{Gamma-low-T}. 
    The latter expresses the $T$-dependence of $\Gamma_{1\to 0}$ near the resonance into a universal form, $G(T / T_*)$ at $\omega_{10}(\Phi) > \delta\Delta$ and $G(T / T_*) e^{-T_* / T}$ at $\omega_{10}(\Phi) < \delta\Delta$, and uses definition $T_*=|\delta\Delta-\omega_{10}|$ which extends the one introduced in the previous sections.
    }
\end{figure}
We start our analysis by considering quasiparticle-induced transitions between two transmon states, $i \neq f$. 
The corresponding transition rate can be obtained from the Fermi's golden rule treating the tunneling Hamiltonian \eqref{H-tunnel} as a perturbation\cite{Catelani-qp}: 
\be \label{Gamma-through-S}
    \Gamma_{i\to f} = \sum_{j=1,2} \big| \langle f|\sin \frac{\hat\varphi_j}{2}|i\rangle \big|^2 \ E_{J,j} \tilde{S}_{qp}(\omega_{if})
\ee
Here $\hat\varphi_j$ are the superconducting phase operators in each junction [$\hat\varphi_1 = \hat\phi - 2\pi \Phi / \Phi_0$ and $\hat\varphi_2 = \hat\phi$, see Eq.~\eqref{H-split-transmon}] and $\omega_{if}$ is the energy difference between the two qubit states; we factored out the Josephson energy from the expression for the structure factor as given by Eqs.~\eqref{S-qp-unexpanded-integrals} and \eqref{S-qp-result}, $\tilde S_{qp} = S_{qp} / E_{J}$.
In the following, we focus on the transmon states $|0\rangle$ and $|1\rangle$.

In the limit $E_J \gg E_C$, the matrix elements in \eqref{Gamma-through-S} can be evaluated using the harmonic oscillator approximation \cite{Catelani-qp}.
\begin{align} \label{matrix-elements-off-diag}
    & \big| \langle 0|\sin \frac{\hat\varphi_j}{2}|1\rangle \big|^2 = \big| \langle 1|\sin \frac{\hat\varphi_j}{2}|0\rangle \big|^2 
    \\ \nonumber
    & \qquad \qquad \qquad = \frac{E_C}{\omega_{10}(\Phi)} \frac{1 + \cos (\pi \Phi/\Phi_0 \pm \vartheta)}{2}.
\end{align}
Substituting expressions for the structure factor Eq.~\eqref{S-qp-result} and matrix element Eq.~\eqref{matrix-elements-off-diag} into the transition rate, Eq.~\eqref{Gamma-through-S}, and expressing $f_0$ through $x_{qp}$ using Eq.~\eqref{f0-via-xqp}, we get the expression for the downward transition rate:
\begin{widetext}
\begin{align}
    \Gamma_{1\to 0} = \frac{ x_{qp}\zeta(T)}{2\pi} 
    \frac{\omega_{10}(0)^2 + \omega_{10}(\Phi)^2}{\omega_{10}(\Phi)}
    \sqrt\frac{\Delta}{2\pi T} 
\Bigg[ & H\left(\frac{|\omega_{10}(\Phi)-\delta\Delta|}{T}\right) \left(\theta_{\omega_{10}(\Phi)-\delta\Delta}+e^{-(\delta\Delta-\omega_{10}(\Phi))/T}\theta_{\delta\Delta-\omega_{10}(\Phi)}\right)
     \nonumber \\ & 
     + e^{-\delta\Delta/T}H\left(\frac{\omega_{10}(\Phi)+\delta\Delta}{T}\right) \Bigg], 
     \label{Gamma-downward}
\end{align}
\end{widetext}
where the function $H(x)$ is defined above Eq.~\eqref{H-asymptotics}. We plot $\Gamma_{1\to 0}$ as a function of temperature in Fig.~\ref{fig:Gamma-10}. 
The opposite, upward in energy transition is given by:
\be \label{Gamma-upward}
    \Gamma_{0\to 1} = \Gamma_{1\to0} \ e^{-\omega_{10}(\Phi)/T} .
\ee
Though transmon is usually operated in the computational basis of states $|0\rangle$ and $|1\rangle$, we note in passing that transition rates for other pairs of adjacent states $n$ and $n-1$ can be obtained from
Eqs.~\eqref{Gamma-downward} and \eqref{Gamma-upward} by multiplying by $n$ and replacing the frequency $\omega_{10}(\Phi)$ with $\omega_{n,n-1}(\Phi)$.

In the limit $T \ll \min(\delta\Delta, \omega_{10})$, Eq.~\eqref{Gamma-downward} for the transition rate simplifies to
\begin{align} \label{Gamma-low-T}
    & \Gamma_{1\to 0} = \frac{ x_{qp}\zeta(T)}{2\pi} \sqrt\frac{\Delta}{2\pi T_*} 
    \frac{\omega_{10}(0)^2 + \omega_{10}(\Phi)^2}{\omega_{10}(\Phi)}
    \\ \nonumber
    &  \ \ \times G\left(\frac{T}{T_*}\right)  \left(\theta_{\omega_{10}(\Phi)-\delta\Delta} + \theta_{\delta\Delta-\omega_{10}(\Phi)} e^{-T_*/T} \right),
\end{align}
where $T_* = |\delta\Delta - \omega_{10}(\Phi)|$ and $G(\tau) = H(1/\tau) / \sqrt{\pi \tau} = e^{1/(2\tau)} K_0(1/(2\tau)) / \sqrt{\pi \tau}$.
The rate $\Gamma_{1\to 0}$ has a logarithmic singularity at $\omega_{10}(\Phi) = \delta\Delta$ due to the singular density of states at the gap edge, as discussed at the end of Section \ref{sec:admittance-Re}. 
In the vicinity of the resonance, $|\delta\Delta - \omega_{10}(\Phi)| \ll \delta\Delta, \omega_{10}(\Phi)$, the transition rate is resonantly enhanced by the $\sqrt{\Delta / (2\pi T_*)}$ factor and varies on the new temperature scale $T_*$, see Fig.~\ref{fig:Gamma-10}. At $T \sim T_*$, this enhancement happens on both sides of the resonance, unlike the frequency shift in Eq.~\eqref{low-T-asymptotics}, which is enhanced only on one side, $\delta\Delta > \omega_{10}(\Phi)$. We observe also that at $\omega_{10}$ slightly exceeding $\delta\Delta$ the relaxation rate drops down with the increase of temperature (see the $\omega_{10}/\delta\Delta=1.05$ curve in Fig.~\ref{fig:Gamma-10} ). This is the result of quasiparticle distribution broadening, which makes the resonant enhancement weaker.

\subsection{Quasi-elastic parity switching rates}
\label{sec:quasi-elastic}

Here we calculate the parity switching rates occurring {\it} without the change of a transmon state. In this case, the energy separation $\omega_{eo}$ between the even/odd states is exponentially small at $E_J \gg E_C$. Therefore, it is typically smaller than the quasiparticle energy $\delta E$, making $\omega_{eo}$ negligible. For that reason, the transition rates from the even to the odd state and vice versa are equal, and we denote them as $\Gamma_{i, e \leftrightarrow o}$, with $i$ being the transmon level index.
The expression for the parity switching rate reads\footnote{We do not include the cross-term in \eqref{Gamma-eo-initial-expression} because it is important only in the crossover region, see Eq.~\eqref{Phi-star}.}:
\begin{align}
    \label{Gamma-eo-initial-expression}
    \Gamma_{i, e \leftrightarrow o}  = & \sum_{j=1,2}\big| \langle i|\sin \frac{\hat\varphi_j}{2}|i\rangle \big|^2 \ E_{J,j} \tilde{S}_{qp}(0)
    \\ \nonumber
    & \quad + \big| \langle i|\cos \frac{\hat\varphi_j}{2}|i\rangle \big|^2 \ E_{J,j} \tilde{S}_{qp}^{(1)}.
\end{align}
Here the first term is a counterpart of Eq.~\eqref{Gamma-through-S} for the $i\to i$ transition, where $\tilde{S}_{qp}$ is evaluated at zero frequency since $\omega_{eo}$ is neglected. The second term arises from the subleading-order expansion of the Bogoliubov amplitudes $u$ and $v$ in the tunneling Hamiltonian and is included because the first term vanishes at $\Phi = 0$, as shown below. It gives the subleading part of the spectral density $\tilde S_{qp}^{(1)}$:
\begin{align}
    \label{Seo-through-integrals}
    \tilde S_{qp}^{(1)} &= \frac{8
    }{\pi} \int_0^{\infty} d\varepsilon \frac{f_{\Delta_L + \varepsilon} \theta_{\varepsilon + \Delta_L - \Delta_R}}{\sqrt{\varepsilon(\varepsilon + \Delta_L - \Delta_R)}} 
    \\ \nonumber
    & \times \left( \frac{\sqrt{\varepsilon} + \sqrt{\varepsilon + \Delta_L - \Delta_R}}{\sqrt{2\Delta}} \right)^2
    + \text{L $\leftrightarrow$ R},
\end{align}
where $f^2$ terms have been neglected since we assume $x_{qp} \ll 1$.

The required matrix elements in the limit $E_J \gg E_C$ are:
\begin{align} \label{matrix-elements-eo}
    \big| \langle i|\sin \frac{\hat\varphi_j}{2}|i\rangle \big|^2 & =  \frac{1 - \cos (\pi \Phi/\Phi_0 \pm \vartheta)}{2},
    \\ \nonumber
    \big| \langle i|\cos \frac{\hat\varphi_j}{2}|i\rangle \big|^2 & =  \frac{1 + \cos (\pi \Phi/\Phi_0 \pm \vartheta)}{2}.
\end{align}
Then, evaluating Eq.~\eqref{Gamma-eo-initial-expression} with the quasi-thermal distribution function \eqref{thermal-f}, expressing the result through $x_{qp}$ according to Eq.~\eqref{f0-via-xqp}, and substituting Eq.~\eqref{vartheta-expression} for $\vartheta$, we arrive at the result
\begin{align} \label{parity-switch-result}
    & \Gamma_{i, e \leftrightarrow o} =
    \frac{ x_{qp}\zeta(T)}{\pi} 
    \frac{\omega_{10}(\Phi)}{E_C}
    \sqrt\frac{\Delta}{2\pi T}  e^{-\delta\Delta/T} 
    \\ \nonumber & \quad
 \times\Bigg\lbrace \frac{\omega_{10}(0)^2 - \omega_{10}(\Phi)^2}{\omega_{10}(\Phi)} H\left(\frac{\delta\Delta}{T}\right)
    \\ \nonumber &
    \quad + \frac{4T}{\Delta} \frac{\omega_{10}(0)^2 + \omega_{10}(\Phi)^2}{\omega_{10}(\Phi)} 
\left[ 1 + \frac{\delta\Delta}{2T} e^{\frac{\delta\Delta}{2T}} K_1\left( \frac{\delta\Delta}{2T}\right) \right] \Bigg\rbrace .
\end{align}
In the experimentally relevant regime $T \lesssim\delta\Delta$, the rate $\Gamma_{i, e \leftrightarrow o}$ grows with temperature, starting with the exponentially suppressed value at small temperatures ($T \ll \delta\Delta$):
\begin{align} \label{parity-switch-low-T}
    & \Gamma_{i, e \leftrightarrow o} =
    \frac{ x_{qp}\zeta(T)}{\sqrt{\pi}} 
    \frac{\omega_{10}(\Phi)}{E_C} \sqrt\frac{\Delta}{2\pi \delta\Delta}  e^{-\delta\Delta/T} 
    \\ \nonumber & \quad
 \times\Bigg\lbrace \frac{\omega_{10}(0)^2 - \omega_{10}(\Phi)^2}{\omega_{10}(\Phi)} 
 + \frac{2\delta\Delta}{\Delta} \frac{\omega_{10}(0)^2 + \omega_{10}(\Phi)^2}{\omega_{10}(\Phi)} 
  \Bigg\rbrace .
\end{align}
The two additive terms here compete with each other, and the competition outcome depends on the flux $\Phi$.
In the experimentally relevant regime, they become comparable at $\Phi\sim\Phi_*$,
\be \label{Phi-star}
    \Phi_* = \Phi_0 \sqrt{\frac{ \max(\delta\Delta, T)}{\Delta}} \frac{\omega_{10}(0)}{\sqrt{[\omega_{10}(0)]^2 - [\omega_{10}(\pi)]^2}};
\ee
this estimate is valid as long as $\Phi_* \ll 1$. 

In the regime of high temperatures,
$T \gg \delta\Delta$, the dependence coincides with the result for a symmetric junction\cite{Catelani-parity-switching, Connolly-nonequilibrium-density}:
\begin{align} 
    & \Gamma_{i, e \leftrightarrow o} =
    \frac{ x_{qp}}{\pi} \frac{\omega_{10}(\Phi)}{E_C}
    \sqrt\frac{\Delta}{2\pi T}  \Bigg\lbrace 
     \frac{\omega_{10}(0)^2 - \omega_{10}(\Phi)^2}{\omega_{10}(\Phi)} \log \frac{T}{\delta\Delta}
    \nonumber \\  &
    \qquad \quad \ \ + \frac{8T}{\Delta} \frac{\omega_{10}(0)^2 + \omega_{10}(\Phi)^2}{\omega_{10}(\Phi)} 
     \Bigg\rbrace.
     \label{parity-switch-high-T}
\end{align}
We see that the first term decreases after reaching a maximum, which can be estimated from Eq.~\eqref{parity-switch-result} to occur at $T \approx 5.1 \delta\Delta$. The second term continues to increase monotonically, further enhancing the parity switching at elevated temperatures. Lastly, we mention that there is no resonance at $\omega_{10}=\delta\Delta$ in the quasi-elastic parity switching rates, as those do not involve transitions between the transmon states. 

\section{Discussion}
\label{sec:discussion}

The goal of this work is to develop a theory facilitating the use of a gap-engineered transmon for studies of the energy distribution of superconducting quasiparticles. Their low density renders ineffective the tunneling spectroscopy approach\cite{Devoret-qpEnergyDistributions}, as it lacks the needed sensitivity. In addition to the desired sensitivity, microwave spectroscopy has far superior resolution, providing access to the low-energy end of the quasiparticle distribution function. This is why we focused on deriving the quasiparticle-induced qubit frequency shift and relaxation rate at small $\delta\Delta-\omega_{10}$. The frequency $\omega_{10}$ of a split-transmon device can be tuned close to the gap difference $\delta\Delta$. The spectroscopic action of the device is allowed by the singular quasiparticle  density of states at the gap edges of the junctions leads. Measurement of the frequency shift $\delta\omega_{10}$, see Section~\ref{sec:frequency-shift}, and the relaxation rate $\Gamma_{1\to 0}$, see Section~\ref{sec:energyrelaxation}, provides two complementary methods to study the quasi-particle distribution function. The quasi-elastic transition rates free of resonances, see Section~\ref{sec:quasi-elastic}, may give a useful reference point for the rates.

One of the outstanding questions of the quasiparticle dynamics   is the energy dependence of their relaxation rate~\cite{CatelaniGlazman-qp-SciPost, SavichGlazmanKamenev, SkvortsovStepanov-GranularTheory}. The use of Eqs.~(\ref{low-T-asymptotics}) and (\ref{Gamma-low-T}) along with the measurements of the type described in Ref.~\onlinecite{Vlad-new-ErrorBursts} may allow accessing it down to the lowest energies.

Quasiparticle temperature is an accepted proxy for their typical energy~\cite{Connolly-nonequilibrium-density}. The theoretical question regarding the detailed form of the distribution function has also recently attracted attention; see, {\sl e.g.}, Ref.~\onlinecite{Basko-noneq-phonons}. To address this more nuanced question with the same measurement techniques~\cite{Vlad-new-ErrorBursts}, one may use the integral representations, Eqs.~\eqref{Sqp-through-integrals} and \eqref{Im-Y-general}, relating measurable quantities to the quasiparticle energy distribution function.

We thank M. H. Devoret, V. D. Kurilovich, A. Opremchak, T. Connolly, and R. J. Schoelkopf for discussions. This research was sponsored by the Army Research Office (ARO) under grants no. W911NF-22-1-0053 and W911NF-23-1-0051, and by the Office of Naval Research (ONR) under award number
N00014-22-1-2764. F. J. M. acknowledges support from the Spanish Ministry of Universities (FPU20/01871, EST24/00678).

\bibliography{asymmetric-qp}

\appendix

\begin{widetext}

\section{Kubo formalism derivation}
\label{sec:Kubo-derivation}

In this section, we derive the full expression for the admittance of an asymmetric Josephson junction for arbitrary ratios $\delta E / \Delta$, $\omega / \Delta$, and $\delta\Delta / \Delta$. We use the Kubo formalism and obtain both real and imaginary parts  without relying on the Kramers–Kronig formula. This derivation formally reproduces the $1/(i\omega)$ contribution from \eqref{Y-three-terms}.

We start by rewriting the tunneling Hamiltonian \eqref{H-tunnel} 
\be
\hat{H}_{T} = e^{i(\varphi_{L}(t)-\varphi_{R}(t))/2}\hat{A}+e^{-i(\varphi_{L}(t)-\varphi_{R}(t))/2}\hat{A}^{\dagger}
\ee
in terms of the operator
\be
	\hat{A}=\sum_{n_{L}n_{R}\sigma}t_{n_{L}n_{R}}c_{n_{L}\sigma}^{\dagger}c_{n_{R}\sigma} .
\ee
We define $\varphi_{t} = \varphi_{L}(t)-\varphi_{R}(t)$ and write down expression for the current using $\hat{I}=-(2e/\hbar)\partial\hat{H}_{T}/\partial \varphi $:
\be
	\hat{I} = \frac{ie}{\hbar} e^{i \varphi_t /2} \hat{A} + \text{h.c.}
\ee
Applying the Kubo formula, we obtain the linear response to the voltage represented by the phase variation \eqref{phi-perturbation}:
\be \label{Kubo-for-current}
	\left\langle \hat{I}\right\rangle =\frac{2e}{\hbar^{2}}\int_{-\infty}^{t}dt_{1} \Re \left\lbrace \left\langle [A^{I}(t),A^{I}(t_{1})]\right\rangle _{0}e^{i\varphi_{t}/2}e^{i\varphi_{t_{1}}/2}+\left\langle [A^{I}(t),A^{I\dagger}(t_{1})]\right\rangle _{0}e^{i\varphi_{t}/2}e^{-i\varphi_{t_{1}}/2}\right\rbrace.
\ee
To proceed, we simplify the averages
\begin{align}
    \left\langle [A^{I}(t),A^{I}(t_{1})]\right\rangle &= \sum_{n_{L}n_{R}\sigma}t_{n_{L}n_{R}}^{2}\ \Big(-\left\langle c_{n_{L}\sigma}^{\dagger}(t)c_{n_{L},-\sigma}^{\dagger}(t_{1})\right\rangle _{0}\left\langle c_{n_{R}\sigma}(t)c_{n_{R},-\sigma}(t_{1})\right\rangle _{0} 
    \\ \nonumber
    & \qquad \qquad \qquad \qquad \qquad \qquad \qquad \qquad \qquad     +\left\langle c_{n_{L},-\sigma}^{\dagger}(t_{1})c_{n_{L}\sigma}^{\dagger}(t)\right\rangle _{0}\left\langle c_{n_{R},-\sigma}(t_{1})c_{n_{R}\sigma}(t)\right\rangle _{0}\Big), 
    \\ \nonumber
    \left\langle [A^{I}(t),A^{I\dagger}(t_{1})]\right\rangle  & = \sum_{n_{L}n_{R}\sigma}|t_{n_{L}n_{R}}|^{2}\left(\left\langle c_{n_{L}\sigma}^{\dagger}(t)c_{n_{L}\sigma}(t_{1})\right\rangle \left\langle c_{n_{R}\sigma}(t)c_{n_{R}\sigma}^{\dagger}(t_{1})\right\rangle -\left\langle c_{n_{R}\sigma}^{\dagger}(t_{1})c_{n_{R}\sigma}(t)\right\rangle \left\langle c_{n_{L}\sigma}(t_{1})c_{n_{L}\sigma}^{\dagger}(t)\right\rangle \right),
\end{align}
and express them in terms of Bogoliubov quasiparticles substituting Eqs.~\eqref{Bogoliubov-transform} and \eqref{uv-expressions}:
\begin{align}
	\left\langle c_{n\sigma}^{\dagger}(t)c_{n',-\sigma'}^{\dagger}(t_{1})\right\rangle & = -\sigma u_{n}v_{n}\delta_{n,n'}\delta_{\sigma,\sigma'}\left(\left\langle \gamma_{n,\sigma}^{\dagger}(t)\gamma_{n,\sigma}(t_{1})\right\rangle _{0}-\left\langle \gamma_{n,-\sigma}(t)\gamma_{n,-\sigma}^{\dagger}(t_{1})\right\rangle _{0}\right)
	\nonumber \\
	 &= -\sigma u_{n}v_{n}\left(-e^{i\tau\epsilon_{n}}+\left\langle \gamma_{n,\sigma}^{\dagger}\gamma_{n,\sigma}\right\rangle _{0}2\cos\tau\epsilon_{n}\right)\delta_{n,n'}\delta_{\sigma,\sigma'},
\end{align}
where $\tau = t_1 - t$ and we used the fact that Bogoliubov quasiparticles are eigenstates of the mean-field Hamiltonian and acquire a simple exponential factor $e^{it \epsilon_n}$ upon time evolution. 
For the conjugate expression,
\be
	\left\langle c_{n\sigma}(t)c_{n',-\sigma'}(t_{1})\right\rangle =\left\langle c_{n',-\sigma'}^{\dagger}(t_{1})c_{n\sigma}^{\dagger}(t)\right\rangle ^{*}=\sigma u_{n}v_{n}\left(-e^{i\tau\epsilon_{n}}+\left\langle \gamma_{n,\sigma}^{\dagger}\gamma_{n,\sigma}\right\rangle _{0}2\cos\tau\epsilon_{n}\right)\delta_{n,n'}\delta_{\sigma,\sigma'}.
\ee
Similarly, for normal averages,
\begin{align}
	\left\langle c_{n\sigma}^{\dagger}(t)c_{n',\sigma'}(t_{1})\right\rangle &= \delta_{n,n'}\delta_{\sigma,\sigma'}\left(u_{n}u_{n'}\left\langle \gamma_{n,\sigma}^{\dagger}(t)\gamma_{n',\sigma'}(t_{1})\right\rangle _{0}+v_{n}v_{n'}\left\langle \gamma_{n,-\sigma}(t)\gamma_{n',-\sigma'}^{\dagger}(t_{1})\right\rangle _{0}\right) \\
	& = \delta_{n,n'}\delta_{\sigma,\sigma'}\left(v_{n}^{2}e^{i\epsilon_{n}\tau}+\left[(u_{n}^{2}-v_{n}^{2})\cos\tau\epsilon_{n}-i\sin\tau\epsilon_{n}\right]\left\langle \gamma_{n,\sigma}^{\dagger}\gamma_{n,\sigma}\right\rangle \right) 
\end{align}
and
\begin{align}
	\left\langle c_{n\sigma}(t)c_{n',\sigma'}^{\dagger}(t_{1})\right\rangle  = \delta_{n,n'}\delta_{\sigma,\sigma'}\left(u_{n}^{2}e^{i\epsilon_{n}\tau}+\left[(-u_{n}^{2}+v_{n}^{2})\cos\tau\epsilon_{n}+i\sin\tau\epsilon_{n}\right]\left\langle \gamma_{n,\sigma}^{\dagger}\gamma_{n,\sigma}\right\rangle \right). 
\end{align}

After some algebra,
\begin{subequations}
\label{Kubo-commutators-result}
\begin{align}
	\left\langle [A^{I}(t),A^{I}(t_{1})]\right\rangle _{0} &= \sum_{n_{L}n_{R}\sigma}t_{n_{L}n_{R}}^{2}u_{n_{L}}v_{n_{L}}u_{n_{R}}v_{n_{R}}\cdot \\ \nonumber
	& \Bigg[2i\sin\tau(\epsilon_{n_{L}}+\epsilon_{n_{R}})-4i\sin\tau\epsilon_{n_{L}}\cos\tau\epsilon_{n_{R}}\left\langle \gamma_{n_{R},\sigma}^{\dagger}\gamma_{n_{R},\sigma}\right\rangle _{0}-4i\sin\tau\epsilon_{n_{R}}\cos\tau\epsilon_{n_{L}}\left\langle \gamma_{n_{L},\sigma}^{\dagger}\gamma_{n_{L},\sigma}\right\rangle _{0}\Bigg],
	\\
	\left\langle [A^{I}(t),A^{I\dagger}(t_{1})]\right\rangle _{0} &=\sum_{n_{L}n_{R}\sigma}|t_{n_{L}n_{R}}|^{2} \Bigg[ u_{n_{R}}^{2}v_{n_{L}}^{2}e^{i\tau(\epsilon_{L}+\epsilon_{R})}-u_{n_{L}}^{2}v_{n_{R}}^{2}e^{-i\tau(\epsilon_{L}+\epsilon_{R})} + \\ \nonumber
	& +\left\langle \gamma_{n_{L},\sigma}^{\dagger}\gamma_{n_{L},\sigma}\right\rangle \left(u_{n_{L}}^{2}u_{n_{R}}^{2}e^{-i\tau(\epsilon_{L}-\epsilon_{R})}-u_{n_{R}}^{2}v_{n_{L}}^{2}e^{i\tau(\epsilon_{L}+\epsilon_{R})}+u_{n_{L}}^{2}v_{n_{R}}^{2}e^{-i\tau(\epsilon_{L}+\epsilon_{R})}-v_{n_{L}}^{2}v_{n_{R}}^{2}e^{i\tau(\epsilon_{L}-\epsilon_{R})}\right) \\ \nonumber
	& +\left\langle \gamma_{n_{R},\sigma}^{\dagger}\gamma_{n_{R},\sigma}\right\rangle \left(-u_{n_{L}}^{2}u_{n_{R}}^{2}e^{-i\tau(\epsilon_{L}-\epsilon_{R})}-u_{n_{R}}^{2}v_{n_{L}}^{2}e^{i\tau(\epsilon_{L}+\epsilon_{R})}+u_{n_{L}}^{2}v_{n_{R}}^{2}e^{-i\tau(\epsilon_{L}+\epsilon_{R})}+v_{n_{L}}^{2}v_{n_{R}}^{2}e^{i\tau(\epsilon_{L}-\epsilon_{R})}\right).
\end{align}
\end{subequations}

Now substituting \eqref{Kubo-commutators-result} into \eqref{Kubo-for-current} and taking the integral over time, we obtain the expression for the current in the system. We are assuming a common distribution function in the leads $f_{\epsilon}$ as introduced in the main text. For the dc Josephson current, the calculation yields the well-known expression \eqref{LJfull-expression}. For the finite-frequency admittance, we substitute \eqref{phi-perturbation} and expand to the linear order in the voltage amplitude $V$. Adding a regularizing factor $e^{0^{+} t}$ and doing some algebra, we obtain the full expressions for the admittance. We present its real part first:
\begin{align}
    \Re Y_{\text{full}}  & = 
    2 G_N \int_{\epsilon_L}^{\infty}d\epsilon_L \int_{\epsilon_R}^{\infty}d\epsilon_R \frac{\epsilon_L \epsilon_R}{\sqrt{(\epsilon_L^2 - \Delta_L^2)(\epsilon_R^2 - \Delta_R^2)}}
      \nonumber \\ 
	& \quad \times \Bigg[ \left(u_{L}^{2}u_{R}^{2}+v_{L}^{2}v_{R}^{2}+2u_{L}u_{R}v_{L}v_{R}\cos\varphi_{0}\right)\left(f_{R}-f_{L}\right)\frac{\delta(\epsilon_{L}-	\epsilon_{R}-\omega)-\delta(\epsilon_{L}-\epsilon_{R}+\omega)}{\omega}  \\  \nonumber
	& \quad \quad \ \ \ + \left(u_{R}^{2}v_{L}^{2}+u_{L}^{2}v_{R}^{2}-2u_{L}u_{R}v_{L}v_{R}\cos\varphi_{0}\right)\left(1-f_{R}-f_{L}\right)\frac{\delta(\epsilon_{L}+\epsilon_{R}-\omega)-\delta(\epsilon_{L}+\epsilon_{R}+\omega)}{\omega}\Bigg].
\end{align}
Substituting Bogoliubov amplitudes from Eq.~\eqref{uv-expressions}, we get
\begin{align}
	\label{ReY-full}
    \Re Y_{\text{full}}  & = 
    G_N \int_{\epsilon_L}^{\infty}d\epsilon_L \int_{\epsilon_R}^{\infty}d\epsilon_R \frac{\epsilon_L \epsilon_R}{\sqrt{(\epsilon_L^2 - \Delta_L^2)(\epsilon_R^2 - \Delta_R^2)}}
      \nonumber \\ 
	& \quad \times \left[
    \left(1 + \frac{\sqrt{\epsilon_L^2 - \Delta_L^2}\sqrt{\epsilon_R^2 - \Delta_R^2}}{\epsilon_L \epsilon_R} + \frac{\Delta_L \Delta_R \cos \varphi_0}{\epsilon_L \epsilon_R} \right)
 \left(f_{R}-f_{L}\right)\frac{\delta(\epsilon_{L}-	\epsilon_{R}-\omega)-\delta(\epsilon_{L}-\epsilon_{R}+\omega)}{\omega}\right.  \\  \nonumber
	& \quad \quad \ \ \ + \left. \left(1 + \frac{\sqrt{\epsilon_L^2 - \Delta_L^2}\sqrt{\epsilon_R^2 - \Delta_R^2}}{\epsilon_L \epsilon_R} - \frac{\Delta_L \Delta_R \cos \varphi_0}{\epsilon_L \epsilon_R} \right) \left(1-f_{R}-f_{L}\right)\frac{\delta(\epsilon_{L}+\epsilon_{R}-\omega)-\delta(\epsilon_{L}+\epsilon_{R}+\omega)}{\omega}\right].
\end{align}
The full expression, including both real and imaginary parts reads
\begin{align}
	\label{Y-full-appendix} & Y_{\text{full}} = \left( \frac{1}{i \omega L_J} + \frac{1}{i \omega L_J^{qp}} \right) \cos \varphi_0 + \frac{G_N}{\pi} \int_{\epsilon_L}^{\infty}d\epsilon_L \int_{\epsilon_R}^{\infty}d\epsilon_R \frac{\epsilon_L \epsilon_R}{\sqrt{(\epsilon_L^2 - \Delta_L^2)(\epsilon_R^2 - \Delta_R^2)}}  \\ \nonumber
    & \quad \times \left[
    \left(1 + \frac{\sqrt{\epsilon_L^2 - \Delta_L^2}\sqrt{\epsilon_R^2 - \Delta_R^2}}{\epsilon_L \epsilon_R} + \frac{\Delta_L \Delta_R \cos \varphi_0}{\epsilon_L \epsilon_R} \right)\left(f_{R}-f_{L}\right)\frac{1}{\epsilon_{L}-\epsilon_{R}}\left(\frac{i}{\omega-\epsilon_{L}+\epsilon_{R}+i0^{+}}+\frac{i}{\omega+\epsilon_{L}-\epsilon_{R}+i0^{+}}\right)\right. \\ \nonumber
    & \quad + \left. \left(1 + \frac{\sqrt{\epsilon_L^2 - \Delta_L^2}\sqrt{\epsilon_R^2 - \Delta_R^2}}{\epsilon_L \epsilon_R} - \frac{\Delta_L \Delta_R \cos \varphi_0}{\epsilon_L \epsilon_R} \right)\left(1-f_{R}-f_{L}\right)\frac{1}{\epsilon_{L}+\epsilon_{R}}\left(\frac{i}{\omega-\epsilon_{L}-\epsilon_{R}+i0^{+}}+\frac{i}{\omega+\epsilon_{L}+\epsilon_{R}+i0^{+}}\right)\right],
\end{align}
where $L_J$ and $L_J^{qp}$ are given by Eqs.~\eqref{LJ-asymmetric} and \eqref{LJ-qp}, correspondingly. 

In the main text, we were interested in the regime $\omega, \delta \Delta, \delta E \ll \Delta_{L}, \Delta_{R}$, when the second line within the square brackets in \eqref{ReY-full} and \eqref{Y-full-appendix} is negligible. Moreover, in that case the Bogoliubov amplitudes $u_n$, $v_n$ can be replaced by there value at $\epsilon_n = \Delta_n$, which is $1/\sqrt{2}$. In that way, Eqs. \eqref{ReY-through-ReYtilde} and \eqref{Im-Y-general} in the main part of this paper were obtained. Should the reader have a need to go beyond the limit $\omega, \delta \Delta, \delta E \ll \Delta_{L}, \Delta_{R}$, the full expressions \eqref{ReY-full} and \eqref{Y-full-appendix} are to be used.

\end{widetext}

\end{document}